\documentstyle[12pt,twoside,fleqn,epsf,espcrc1]{article}
\input psfig
\newcommand{\be}{\begin{eqnarray}}
\newcommand{\ee}{\end{eqnarray}}

 \def\gsim{\mathrel{\raise.3ex\hbox{$>$}\mkern-14mu
	      \lower0.6ex\hbox{$\sim$}}}
 \def\lsim{\mathrel{\raise.3ex\hbox{$<$}\mkern-14mu
	      \lower0.6ex\hbox{$\sim$}}}

 \newcommand{\AmS}{{\protect\the\textfont2
   A\kern-.1667em\lower.5ex\hbox{M}\kern-.125emS}}

 \newcommand{\ga}{\gamma}
 \newcommand{\bk}{{\bf k}}
 
 \newcommand{\psib}{{\bar\psi}}

 \renewcommand{\slash}{\!\!\!\!/\,}
 
 \newcommand{\dsp}{\displaystyle}

 \title{QCD In Extreme Conditions
 \thanks{Lectures given at CRM Summer
 School, June 27-July 10, Banff (Alberta), Canada.}}

 \author{Frank Wilczek\address{School of Natural Science, 
	 Institute for Advanced Study\\ 
	 Princeton, NJ 08540 USA}
	 \thanks{Research supported in part by DOE grant
		 DE-FG02-90ER40542. IASSNS-HEP-99-92 e-mail:
	 wilczek@sns.ias.edu}}

\begin{document}

 \maketitle

 \begin{abstract} Recently we have made considerable progress in our
 understanding of the behavior of QCD in extreme conditions of high
 temperature or large baryon number density.  Among the highlights are
 the prediction of a well-characterized true critical point, and the
 discovery that the ground state of three-flavor QCD at asymptotically
 high densities exhibits color-flavor locking.  The critical point
 occurs at the unique temperature and density where a sharp
 distinction between an ionized plasma of quarks and gluons and the
 hadronic phase first appears.  It appears to be accessible both to
 numerical and to laboratory experiments.  Color-flavor locking
 provides a calculable, weak-coupling realization of confinement and
 chiral symmetry breaking.  It also provides a microscopic realization
 of Han-Nambu charge assignments for quark quasiparticles, and of
 Yang-Mills theory for the physical vector mesons.  Here I provide a
 self-contained introduction to these developments.

 \end{abstract}

\section{Introduction}

In some ways, QCD is a mature subject.  Its principles are precisely
defined, and they have been extensively confirmed by experiment.  QCD
specifies unambiguous algorithms, capable of transmission to a Turing
machine, that supply the answer to any physically meaningful question
within its domain -- any question, that is, about the strong
interaction.  I believe there is very little chance that the
foundational equations of QCD will require significant revision in the
foreseeable future.  Indeed, as we shall soon review, these equations
are deeply rooted in profound concepts of symmetry and local quantum
field theory, which lead to them uniquely.  So one cannot revise the
equations without undermining these concepts.

Granting that the foundations are secure, we have the task -- which is
actually a wonderful opportunity -- of building upon them.  Due to the
peculiarities of QCD, this is a particularly interesting and important
challenge.

Interesting, because while the foundational equations are conceptually
simple and mathematically beautiful, they seem at first sight to have
nothing to do with reality.  Notoriously, they refer exclusively to
particles (quarks and gluons) that are not directly observed.  Less
spectacular, but more profound, is the fact that equations exhibit a
host of exact or approximate symmetries that are not apparent in the
world.  One finds that these symmetries are variously hidden: confined
in the case of color, anomalous in the cases of scale invariance and
axial baryon number, spontaneously broken in the case of chirality.
It is fascinating to understand how a theory that superficially
appears to be ``too good for this world'' actually manages to describe
it accurately.  Conversely it is pleasing to realize how the world is,
in this profound and very specific way, simpler and more beautiful
than it at first appears.

Important, because there are potential applications to which the
microscopic theory has not yet rendered justice.  An outstanding,
historic, challenge is to derive the principles of nuclear physics.
At present meaningful connections between the microscopic theory of
QCD and the successful practical theory of atomic nuclei are few and
tenuous, though in principle the former should comprehend the latter.
This seems to be an intrinsically difficult problem, probably at least
as difficult as computing the structure of complex molecules directly
from QED.  In both cases, the questions of interest revolve around
small energy differences induced among valence structures, after
saturation of much larger core forces.  If one starts calculating from
the basic equations, unfocussed, then small inaccuracies in the
calculation of core parameters will blur the distinctions of interest.

A related but simpler class of problems is to calculate the spectrum
of hadrons, their static properties, and a variety of operator matrix
elements that are vital to the planning and interpretation of
experiments.  This is the QCD analogue of atomic physics.  Steady
progress has been made through numerical work, exploiting the full
power of modern computing machines.

Other significant applications appear more accessible to analytic
work, or to a combination of analytics and numerics.   

The behavior of QCD at high temperature and low baryon number density
is central to cosmology.  Indeed, during the first few seconds of the
Big Bang the matter content of the Universe was almost surely
dominated by quark-gluon plasma.  There are also ambitious, extensive
programs planned to probe this regime experimentally.

The behavior of QCD at high baryon number density and (relatively) low
temperature is central to extreme astrophysics -- the description of
neutron star interiors, neutron star collisions, and conditions near
the core of collapsing stars (supernovae, hypernovae).  Also, we might
hope to find -- and will find -- insight into nuclear physics, coming
down from the high-density side.

The special peculiarity of QCD, that its fundamental entities and abundant
symmetries are well hidden in ordinary matter, lends elegance and
focus to the discussion of its behavior in extreme conditions.
Quarks, gluons, and the various symmetries will, in the right
circumstances, come into their own.  By tracing symmetries lost and
found we will be able to distinguish sharply among different phases of
hadronic matter, and to make some remarkably precise predictions about
the transitions between them.

In these 5 lectures I hope to provide a reasonably self-contained
introduction to the study of QCD in extreme conditions.  The first
lecture is a rapid tour of QCD itself.  I've organized it as a
cluster of related stories narrating how apparent symmetries of the
fundamental equations are hidden by characteristic dynamical
mechanisms.  Lectures 2-3 are mainly devoted to QCD at high
temperature, and lectures 4-5 to QCD at high baryon number density.  I
have structured the lectures so that they head toward two climaxes:
the prediction of a true critical point in real QCD, that ought to be
accessible to numerical and laboratory experiments, in Lecture 3, and
the prediction that at asymptotic densities QCD goes over into a
color-flavor locked phase with remarkable properties including fully
calculable realizations of confinement and chiral symmetry breaking,
in Lecture 5.  These are, I believe, remarkable results, and they
bring us to frontiers of current research.

A wide variety of tools will be brought to bear, including three
different renormalization groups (the usual `asymptotic freedom'
version toward bare quarks and gluons at high virtuality, the usual
`Kadanoff-Wilson-Fischer' version toward critical modes at a
second-order phase transition, and the `Landau-Anderson' version
toward quasiparticles near the Fermi surface), perturbative quantum
field theory, effective field theory, instantons, lattice gauge
theory, and BCS pairing theory.  Of course I won't be able to give
full-blown introductions to all these topics here; but I'll try to give
coherent accounts of the concepts and results I actually need, and to
supply appropriate standard references.  In a few months, I hope, a
comprehensive reference will be appearing.

\section{Lecture 1: Symmetry and the Phenomena of QCD}

\subsection{Apparent and Actual Symmetries}

Let me start with a slightly generalized and slightly idealized
Lagrangian for QCD: 
\begin{equation}
\label{masslessL}
{\cal L} ~=~ - {1\over 4 g^2} {\rm tr}~ G^{\mu \nu} G_{\mu \nu} +
\sum^f_{j=1} \bar \psi_j 
(i\!\not\!\!D ) \psi_j
\end{equation}
where 
\begin{equation}
\label{fieldStrength}
G_{\mu \nu} ~=~ \partial_\mu A_\nu - \partial_\nu A_\mu + i [A_\mu , A_\nu ]
\end{equation}
and
\begin{equation} 
\label{covariantD}
D ~=~ \partial_\mu + i A_\mu
\end{equation}
The $A_\mu$ are 3$\times$3 traceless Hermitean matrices.  Each spin-1/2
fermion quark field $\psi_j$ carries a corresponding 3-component
color index.   

Eqn.~\ref{masslessL} is both slightly generalized from real-world QCD, in
that I've allowed for a variable number $f$ of quarks; and slightly
idealized, in that I've set all their masses to zero.  Also, I've set
the $\theta$ parameter to zero, once and for all.  We will have much
occasion to focus on particular numbers and masses of quarks in our
considerations below, but Eqn.~\ref{masslessL}
is a good simple starting point.

Our definition of the gauge potential differs from the more traditional one by
\begin{equation}
\label{normalizeA}
A_\mu^{\rm here} ~=~ g A_\mu^{\rm usual}
\end{equation}
which is why the coupling constant does not appear in the covariant
derivative.  By writing the Lagrangian in this form we make it clear
that $1/g$ is  neither more nor less than a stiffness parameter.  It
informs us how big is the energetic cost to produce curvature in the
gauge field.

The form of
Eqn.~\ref{masslessL} is uniquely fixed by a few abstract postulates of
a very general character.  These are $SU(3)$ gauge symmetry, together
with the general principles of quantum field theory -- special
relativity, quantum mechanics, locality -- and the criterion of
renormalizability.  It is renormalizability that forbids more
complicated terms, such as an anomalous gluomagnetic moment term
$\propto \bar q \sigma^{\mu\nu} G_{\mu \nu} q$.

Later I shall argue that it is very difficult to rigorously insure the
existence of a quantum field theory that is not asymptotically free.
Asymptotic freedom is a stronger requirement than renormalizability,
and can only be achieved in theories containing nonabelian gauge
fields.  Thus one can say, without absurdity, that even our
postulates of gauge symmetry and renormalizability are gratuitous:
both are required for the {\it existence\/} of a relativistic local
quantum field theory.

The apparent symmetry of Eqn.~\ref{masslessL} is:
\begin{equation}
\label{apparentG}
{\cal G}_{\rm apparent} ~=~ SU(3)^c \times SU(f)_L \times SU(f)_R
\times U(1)_B \times U(1)_A \times {\cal R}^+_{\rm scale}~,  
\end{equation}
together of course with Poincare invariance, P, C, and T.  The
factors, in turn, are local color symmetry, the freedom to freely
rotate left-handed quarks among one another, the freedom to freely
rotate right-handed quarks among one another, baryon number (= a
common phase for all quark fields), axial baryon number (= equal and
opposite phases for all left-handed and right-handed quark fields), 
and scale invariance.  

The chiral $SU(f)_L\times SU(f)_R$ symmetries arise because the only
interactions of the quarks, their minimal gauge couplings to gluons,
are universal and helicity conserving.  These chiral symmetries will
be spoiled by non-zero quark mass terms, since such terms connect the
two helicities.  With quarks of non-zero but equal mass one would have
only the diagonal (vector) $SU(f)_{L+R}$ symmetry, while if the quarks
have unequal non-zero masses this breaks up into a product of $U(1)$s.
Thus the choice $m=0$ can be stated as a postulate of enhanced
symmetry.  If the quarks masses are all non-zero, then P and T would
be violated by a non-zero $\theta$ term, unless $\theta = \pi$.  For massless
quarks, all values of $\theta$ are physically equivalent, so we lose
nothing by fixing $\theta = 0$.

One could of course have written these chiral symmetries together with
the two $U(1)$ factors as $U(f)_L\times U(f)_R$, but the unwisdom of
so doing will become apparent momentarily.

Finally the ${\cal R}^+_{\rm scale}$
factor reflects that the only parameter in the theory, $g$, is
dimensionless (in units with $\hbar = c=1$, as usual).  Thus the classical
theory is invariant under a change in the unit used to measure length,
or equivalently (inverse) mass.  Indeed, the action 
$\int d^4x  {\cal L}$ is invariant under the
rescaling
\begin{equation}
x^\mu \rightarrow \lambda x^\mu; A \rightarrow \lambda^{-1} A;  
\psi \rightarrow \lambda^{-1} \psi~. 
\end{equation} 

The actual symmetry of QCD, and of the real world, is quite different
from the apparent one.  It is
\begin{eqnarray}
\label{actualG}
{\cal G}_{\rm actual} 
& = & SU(3)^c \times SU(f)_L  \times SU(f)_R \times Z_A^{f} \times
U(1)_B \nonumber \\ 
&~ & {\rm (asymptotic~ freedom,~  chiral~  anomaly)} \nonumber \\ 
& \rightarrow & SU(3)^c \times SU(f)_{L+R} \times U(1)_B \nonumber \\
&~ & {\rm (chiral~  condensation)} \nonumber \\
& = &  SU(f)_{L+R} \times U(1)_B \nonumber \\
&~ & {\rm (confinement)}. 
\end{eqnarray} 
\noindent Let me explain this cascade of symmetry reductions.

In the first line of Eqn.~\ref{actualG}, 
I've specified the subgroup of ${\cal G}_{\rm apparent}$
which survives quantization.  The $ {\cal R}^+_{\rm scale}$ of
classical scale invariance is entirely lost, and the $U(1)_A$ of axial
baryon number is reduced to its discrete subgroup $Z_A^{f}$.  Both
fall victims to the need to regulate quantum fluctuations of highly
virtual degrees of freedom, as I shall elaborate below.  The breaking
of scale invariance is associated with the running of the effective
coupling, asymptotic freedom, and dimensional transmutation.  The
breaking of axial baryon number is associated with the triangle
anomaly and instantons.  Thus these symmetry removals 
arise from dynamical features of QCD that reflect its
deep structure as a quantum field theory. 

In the second line of Eqn.~\ref{actualG}, 
I've specified the subgroup of the symmetries of the
quantized theory which are also symmetries of the ground state.  This
group is properly smaller, due to spontaneous symmetry breaking.  That
is, the
stable solutions of the equations exhibit less symmetry than the
equations themselves.  Specifically, 
the ground state contains a condensate of
quark-antiquark pairs of opposite handedness, which fills space-time
uniformly.  One cannot rotate
the different helicities components independently while leaving the
condensate invariant.  The lightness of $\pi$ mesons, and much of the
detailed phenomenology of their interactions at low energies, can be
understood as direct consequences of this spontaneous symmetry
breaking.

In the third line of Eqn.~\ref{actualG}, I've acknowledged that local color
gauge symmetry, which is so vital in formulating the theory, is
actually not directly a property of any physical observable.  Indeed,
in constructing the Hilbert space of QCD, one must restrict oneself to
gauge-invariant states.  The auxiliary, extended Hilbert space that we
use in perturbation theory does not have a positive-definite inner
product.  It's haunted by ghosts.  Moreover, unlike the situation for
QED, one discerns in the low-energy physics of QCD no obvious traces
of gauge symmetry.  Specifically, there are no long-range forces, nor
do particles come in color multiplets.  This is the essence of
confinement, a tremendously important but amazingly elusive concept,
as we shall discover repeatedly in these lectures.  (If you look only
at the world, not at a postulated micro-theory, what exactly does
confinement mean?  Don't fall into the trap of saying confinement
means the unobservability of quarks -- the quarks are a theoretical
construct, not something you can observe (that's what you said!).)

Clearly, a major part of understanding QCD must involve understanding how
its many apparent symmetries are lost, or realized in peculiar ways.  The
study of QCD in extreme conditions gives us fruitful new perspectives
on these matters, for we can ask new, very sharp questions: Are
symmetries restored, or lost, in phase transitions? Are they restored
asymptotically?

\bigskip
Now I shall discuss each of the key dynamical phenomena: asymptotic
freedom, confinement, chiral condensation, and chiral anomalies, in
more detail.

\subsection{Asymptotic Freedom}

\subsubsection{Running Coupling}

Running of couplings is a general phenomenon in quantum field theory.
Nominally empty space is full of virtual particle-antiparticle pairs
of all types, and these have dynamical effects.  Put another way,
nominally empty space is a dynamical medium, and we can expect it to
exhibit medium effects including dielectric and paramagnetic behavior,
which amount (in a relativistic theory) 
to charge screening.  In other words, the strength of the
fields produced by a test charge will be modified by vacuum
polarization, so that the {\it effective\/} value of its charge depends on the
distance at which it is measured.

Asymptotic freedom is the special case of running couplings, in which
the effective value of a charge measured to be finite at a given
finite distance,
decreases to zero when measured at very short distances.
Thus asymptotic freedom involves antiscreening.  Antiscreening is somewhat
anti-intuitive, since it is the opposite of what we are accustomed
to in elementary electrodynamics.  Yet there is a fairly simple way to
understand how it might be possible in a {\it relativistic, nonabelian,
gauge theory}:  

\begin{itemize}
\item Because the theory is relativistic, magnetic forces are
just as important as electric ones.  

\item Because it is a gauge theory, it
contains vector mesons, with spin.  

\item Because it is nonabelian these
vector mesons carry charge, and their spins carry magnetic moments.
\end{itemize}

\noindent Now in electrodynamics 
we learn that spin response is paramagnetic --
a spin (elementary magnetic dipole) 
tends to align with an imposed magnetic field, in such a way as
to enhance the field.  This, you'll realize if you think about it a moment, is
antiscreening behavior.  Thus there is a competition between normal
electric screening (together with orbital diamagnetism) and
antiscreening through spin paramagnetism.  For virtual gluons, it
turns out that spin paramagnetism is the dominant effect numerically.

\begin{figure}[htb]
\centerline{\psfig{figure=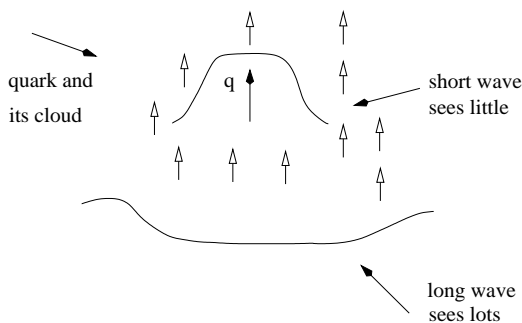,width=7cm}}
\vglue-.4in
\caption[]{Distribution of the color charge in a cloud.}
\label{fig:cloud.eps}
\end{figure}

\begin{figure}[!t]
\centerline{\psfig{figure=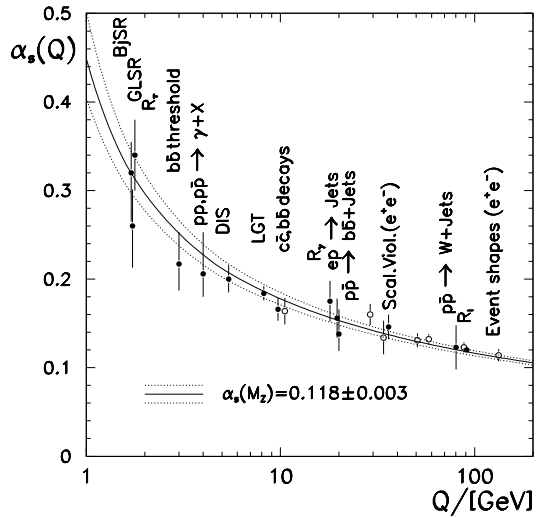,width=7cm}}
\vglue-.4in
\caption[]{Running of the couplings from \cite{Sch}.} 
\label{fig:asdrun.eps}
\end{figure}

When the effective coupling becomes weak, one can calculate the
screening (or antiscreening) behavior in perturbation theory.  In QCD
one finds for the change in effective coupling
\begin{equation}
\label{runningEqn}
{dg(\epsilon) \over d \ln \epsilon} ~=~ \beta_0 g^3 + \beta_1 g^5 + ...  
\end{equation}
with
\begin{eqnarray}
\label{rgCoeffs}
\beta_0 ~& = & ~  {-1\over 16 \pi^2}( 11-{2\over 3} f) \cr
\beta_1 ~& = & ~  ({1\over 16 \pi^2})^2 ( 102 - {38\over 3} f) ,
\end{eqnarray}
where $\epsilon$ is the energy, or equivalently inverse length, scale
at which the effective charge is defined.  Thus if the $f \leq 16$ the
effective coupling decreases to zero as the energy at which it is
measured increases to infinity, which is asymptotic freedom.  Taking
the first term only, we see that the asymptotic behavior is
approximately
\begin{equation}
\label{runningSol}
{1\over g(\epsilon)^2} ~=~ {1\over g(1)^2} - \beta_0  \ln \epsilon.
\end{equation}

\subsubsection{Clouds, Jets, and Experiments}

The asymptotic freedom of QCD can be exploited to simplify the
calculation of many physical processes.  To understand why that is so,
consider the picture of quarks (or gluons) that it suggests.  This is
shown in Figure 1.  The effective color charge of a quark, responsible
for its strong interaction, accumulates over distance.  We should
think of it as a distributed cloud of induced color charge, all of the same
sign, with no singular core.  Now consider how this cloud is seen when
probed at various wavelengths.  If the wavelength is small, the
effective color charge will be small, since only a small portion of
the cloud is sampled.  If the wavelength is large, the effective
charge will be large.  

One can also consider the complementary
particle picture.  If one suddenly imparts a large impulse to a quark
(or gluon), liberating it from its cloud, the resulting object will
have a small effective color charge, and will propagate almost freely,
until it builds up a new cloud.

Of course the electric and weak charges of a quark are not shared by
its color polarization cloud.  These charges remain concentrated.
Thus short-wavelength electromagnetic or weak probes can resolve
pointlike quarks, whose `strong' (color) interactions make only small
corrections to free-particle propagation.  More precisely, it is
corrections that substantially change the energy-momentum of the color
current which are guaranteed to be small.  Big changes in the
energy-momentum can only arise from radiation of gluons having short
wavelength in the frame of the current, but such short-wavelength
gluons see only a small effective color charge, as we've discussed.

So in physical process originating with the pointlike quarks directly
produced by hard electroweak currents, as for instance in $e^+e^-$
annihilation at high energy, one develops jets of rapidly-moving
hadrons following the nominal paths of the original quarks.  Hard
gluon radiation processes do occasionally occur, of course, but at a
small rate, calculable in perturbation theory.  These hard gluons can
(with small probability) induce jets of their own, which themselves
occasionally radiate, and so forth.  The ``antenna pattern'' of jets
-- including the relative probabilities for different topologies
(numbers of jets), the energy and angular distributions for a given
topology, and how all these quantities depend upon the total energy,
can be calculated in exquisite detail.  These predictions, which
reflect in 1-1 fashion the structure of the fundamental interactions
in the theory and the running of the coupling, have been extensively,
and successfully, tested experimentally.

The science of exploiting asymptotic freedom to predict rates for a
wide variety of  hard
processes is highly developed.  Figure 2 displays results of
comparisons between prediction and observation for a wide variety of
experiments, in the form of determinations of the running coupling.
Implicit, of course, is that within any given type of experiment, the
theory gives a successful account of the functional form of dependence
on event topology, energy distribution, angular distribution, ... .  
The Figure
itself demonstrates the success of Eqns.~\ref{runningEqn},\ref{rgCoeffs}, 
suitably
generalized to include the effect of quark masses, as a description of
Nature.

\subsubsection{Dimensional Transmutation and ``Its From Bits''}

Running of the coupling manifestly breaks the classical scale
invariance ${\cal R}^+_{\rm scale}$.  It is an almost inevitable
result of quantum field theory.  Indeed, quantum field theory predicts
as its first consequence the existence of a space-filling medium of
virtual particles, whose polarization makes the effective coupling
strength depend on the distance at which it is measured, or in other
words causes the coupling to run. (There are very special classes of
supersymmetric theories in which different types of virtual particles
give cancelling effects, and the coupling does not run.)

Because its coupling runs, in QCD the analogue of Pauli's question: 
\begin{quote}
``Why is the value of the fine structure constant what it is?'' -- 
\end{quote}

\noindent receives a startling answer: 

\begin{quote}
 ``It's anything you like -- at some scale or other.'' 
\end{quote}

\noindent We can simply declare it to be, say, ${1\over 10}$, thereby
defining the correct distance at which to measure it -- {\it i.e.},
the distance where it is ${1\over 10}$!  This is the phenomenon of
dimensional transmutation.  A dimensionless coupling constant has
been transmuted into a length (or, equivalently, energy) scale.

In fact we can identify a scale explicitly, using Eqn.~\ref{runningEqn}:
\begin{equation}
\label{uvScale}
M ~=~ \lim_{\epsilon \rightarrow \infty}  \epsilon e^{u_{}\over \beta_0}
u^{\beta_1\over \beta_0^2} 
\end{equation}
where
\begin{equation}
\label{uDef}
u ~\equiv~ {1\over g(\epsilon)^2} ~.
\end{equation} 
The form of Eqn.~\ref{runningEqn}
insures that the limit exists and is finite.

Due to its formal scale invariance, QCD with massless quarks appears
naively -- that is, classically -- to be a on-parameter family of
theories, distinguished from one another by the value of the coupling
$g$.  And none of these theories, it appears, defines a scale of
distance.  Through the magic of dimensional transmutation, QCD turns
out instead to be a family of perfectly identical clones, each of
which {\it does\/} define a distance scale.  Indeed, the clones differ
only in the units they employ to measure distance.  This difference in
units enters into comparisons of purely QCD quantities to quantities
outside of QCD, such as the ratio of the proton mass to the electron
mass.  But it does not affect dimensionless quantities within QCD
itself, such as ratios of hadronic masses or of isomer splittings.

Using $\hbar$ and $c$ as units, and {\it no further inputs\/} the
truncated version of QCD including just the $u$ and $d$ quarks, with
their masses set to zero -- what I call ``QCD Lite'' -- accounts
pretty accurately for the low-lying spectrum of non-strange hadrons
(demonstrably), nuclear physics (presumably), and much else besides.
The only numerical inputs to QCD Lite are the number of colors -- 3
(binary 11) -- and the number of quarks --2 (binary 10).  Thus QCD Lite
provides a remarkable realization of Wheeler's slogan ``Getting Its From
Bits''!

\subsubsection{Limitations}

The most straightforward applications of asymptotic freedom are to
processes involving large energy scales only.  This constraint pretty
much limits one to inclusive processes induced by electroweak
currents.  Any external hadron introduces a small energy scale, namely
its mass.  By using clever tricks to isolate calculable subprocesses,
one can calculate much more, as conveyed in Figure 2.  However these
tricks will take you only so far, and there is no easy way to exploit
the small effective coupling at short distances to address truly
low-energy phenomena, or to calculate the spectrum.  If you try to
calculate such things directly, you will encounter infrared
divergences -- so at least the theory is kind enough to warn you off.
Similarly, large temperature or large chemical potential introduce
large ``typical'' energies, but it is not trivial (and usually not
even true) that this fact in itself will allow access, via asymptotic
freedom, to interesting spectral or transport properties.

\subsection{Confinement}
\subsubsection{Brute Facts and Crude Theory}

An aura of mystery still seems to hover around the phenomenon of quark
confinement.  Historically, of course, it was a big surprise in
world-modeling, and posed a major barrier both to the discovery and to
the acceptance first of quarks, and then of modern QCD.  And confinement is a
genuinely profound and subtle dynamical phenomenon, as some of our
later considerations will emphasize.  Concerning the {\it fact\/} that
QCD predicts confinement, however, there is no ambiguity.  Our
knowledge of the properties of the theory has moved far beyond
abstract or hypothetical discussion of this point.  Figure 3 exhibits
the results of some direct calculations of the spectrum starting from
the microscopic theory, with controlled errors, using the techniques
of lattice gauge theory.  There are neither massless flavor-singlet
particles with long-range interactions, particles with quark or gluon
quantum numbers, nor degenerate color multiplets.  In fact the
microscopic theory reproduces the observed spectrum extremely well,
with no gratuitous additions.

Thus confinement is not a {\it practical\/} problem for modern QCD.
Still, one would like to understand more precisely what it is, why it
occurs and, particularly for these lectures, whether in can come
undone in extreme conditions.

\begin{figure}[!t]
\centerline{\psfig{figure=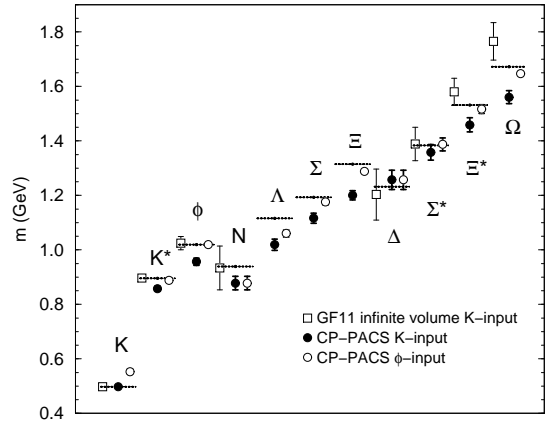,width=7cm}}
\vglue-.4in
\caption[]{Spectrum from lattice gauge theory, from \cite{Burk}.} 
\label{fig:spectrum.eps}
\end{figure}

The simplest heuristic argument for confinement is due to Amati and
Testa.  It is based directly on Eqn.~\ref{masslessL}.  If the coefficient
${1\over g^2}$ of the gauge curvature term is taken to zero, then upon
varying the action with respect to the gauge potential $A_\mu$ we find
that the color current vanishes, including its zero component, the
color charge density -- which is to say, color is confined.  We can
combine this idea with the running of the coupling, to argue (still
more heuristically) that low-frequency modes, associated with large
effective couplings, are confined, while high-frequency modes can be
dynamically active.  This is broadly consistent with the observed
behavior in Nature, that quarks and gluons become visible in hard
processes, but are not accessible to soft probes.

Unfortunately it is a very singular operation to throw away the
highest-derivative terms in any differential equation, because they
will always dominate for sufficiently abrupt variation.  In high
Reynolds number hydrodynamics, one addresses this difficulty
with boundary layer theory.  In QCD, the only method we know to make
the simple Amati-Testa argument the starting point for a systematic
approximation is by employing an artful discretization, lattice gauge theory,
as I'll now discuss.

\subsubsection{Lattice Gauge Theory Basics}

The great virtue of the lattice version of QCD are that it provides an
ultraviolet cutoff and that it allows
a convenient strong-coupling expansion, while
preserving a very large local gauge symmetry.  Its drawbacks are that
it destroys translation and rotation symmetry, that it has an awkward
weak-coupling expansion, and that it mutilates the ultraviolet
behavior of the continuum theory.  But
let's put these worries aside for the moment, and exploit the virtues.
Also let's consider the pure glue theory.  The extension to
quarks will appear in Lecture 2. 

The fundamental operation in gauge theory is parallel transport.  The
basic objects of the theory are 3$\times$3 unitary matrices with unit
determinant.  They live on the oriented links of a cubic hyperlattice
in four dimensions, and implement parallel transport.  Thus the
dynamical variable $U_{r, \hat \mu}$ is a matrix associated with the
link starting from lattice point $r$ and ending at $r + \hat \mu a$,
where $a$ is the lattice spacing.  The matrix associated with the same
link oriented in the opposite direction is the inverse, {\it i.e}.
$U_{r + \hat \mu a, -\mu} ~=~ U_{r, \hat \mu}^{-1}$.

If there were an underlying continuum gauge potential
$A$, the parallel transport would be
\begin{equation}
\label{uTransport}
U_{r, \hat \mu}~  \hbox{``}=\hbox{''} ~\bar P \exp i \int^{r+ \hat
\mu a}_r A_\mu dx^\mu~, 
\end{equation}
where $\bar P$ denotes anti-path ordering.  Indeed, this is the solution of
the equation  
\begin{equation}
\nabla_\mu U ~\equiv~ (\partial_\mu + i A_\mu )U ~=~ 0 
\end{equation}
having the indicated endpoints.  In line with this underlying
structure, which we would like to recover in an appropriate limit, 
local gauge
transformations $\Omega (r)$ should act on the lattice sites, and
transform the $U$ matrices according to
\begin{equation}
\label{transformU}
{\tilde U}_{r, \mu} ~=~ \Omega (r) U_{r, \mu} \Omega(r + \hat \mu a )^{-1}~.
\end{equation}
The $\Omega(r)$ are, of course, 3$\times$3 unitary matrices.   

To form interaction terms invariant under Eqn.~\ref{transformU} one must
take the trace of a product of $U$ matrices along a path of links
forming a closed loop.  The simplest possibility is simply the trace
around a plaquette, {\it e.g}.
\begin{equation}
\label{plaqVar}
{\rm tr}~{\it Pl}_{r,\hat x \hat y}~\equiv~ {\rm tr}~
U_{r, \hat x} U_{r + a \hat x, \hat y} U_{r + a \hat x + a \hat y,
-\hat x} U_{r+ a\hat y, -\hat y} ~. 
\end{equation} 
Putting Eqn.~\ref{uTransport}
into Eqn.~\ref{plaqVar}
and  expanding for small
$a$, we find the first non-trivial term  
\begin{equation}
\label{expandPlaq}
{\rm tr}~{\it Pl}_{r,\hat x \hat y} ~ \rightarrow  {\rm tr}~ (1 -
{a^4\over 2} G^{xy}G_{xy} )  
+ O(a^6)~.
\end{equation}
The straightforward verification of Eqn.~\ref{expandPlaq} is quite arduous,
but one can restrict its form {\it a priori\/} by exploiting gauge
invariance, and then evaluate on a simple configuration such as $A_x
~=~ M (y - r_y) $.  Note that a term $\propto a^2 G^{xy}$, which does
appear in the plaquette product, vanishes when we take the trace.

Thus the simplest lattice gauge invariant action we can write, the
Wilson action
\begin{equation}
\label{wilsonS}
S_W ~=~ {1\over 4 g^2} \sum_{\rm plaquettes} (3 - {\rm tr}~{\it Pl}_{\Box})~~ 
\end{equation}
reduces, formally, to the continuum action for very small lattice
spacings.  Hence we are invited to use it, and then attempt to justify the
limit.  (In the spirit of the Jesuit credo ``It is more blessed to ask
forgiveness than permission.'')

To evaluate a correlation function of operators $\langle {\cal O}_1
{\cal O}_2 ... \rangle$, then, we must evaluate the integral
\begin{equation}
\label{latExpV}
{\langle {\cal O}_1 {\cal O}_2 ... \rangle ~=~
\int d{\cal M} \exp (- S_W)
{\cal O}_1 {\cal O}_2 ... \over  \int d {\cal M} \exp (- S_W) }~, 
\end{equation}
where
\begin{equation}
d{\cal M} ~=~ 
{\textstyle\prod}_{{\rm links}~r, \hat \mu}U_{r, \hat \mu}^{-1}
dU_{r, \hat \mu} 
\end{equation}
is the product of invariant integrals over the gauge group.  

\subsubsection{Strong Coupling and Confinement}

The lattice regularization permits one to formulate strong coupling
perturbation theory in a simple, elegant way.  When $g$ gets large, we
can simply expand $e^{-S_W}$ in a power series in ${1\over g^2}$.  The
result is to ``bring down'' activated plaquettes, one for each inverse
power of ${1\over g^2}$.  When we integrate over a link incident on an
activated plaquette, we encounter an extra power of $U$ for that link.

To test for confinement, the traditional method is to study
Wilson-Polyakov loops.  These are simply the traces of products of $U$
matrices around loops, similar to the terms that appear in the action.
But now we want to consider large loops, instead of very small ones.

The motivation for this method
is as follows.  Suppose we put a very heavy quark into
the system.  This will stay at a fixed point in space, so its
world-line will be simply a straight line in the $\hat \tau$
(Euclidian time) direction.  The ${\rm tr}~j\cdot A$ coupling of this
color source will generate a product of $U$ matrices along the links
of its world-line.  So to measure the potential between a heavy quark
and a heavy antiquark at distance $R$ we should measure the energy it
takes to have a line like this and a similar line with $U^{-1}$
matrices a distance $R$ away.  If we allow this configuration to
persist for a long Euclidian time $T$, the cost should go as
$e^{-V(R)T}$.  Now to make the ``measurement'' clean we should
imagine closing up the loop with short segments at the top and bottom.
This corresponds to producing the quark-antiquark pair, letting them
sit separated for a long time, and then annihilating them.  With
$R<<T$, by taking the negative of the log, we will extract the
potential.  In a formula
\begin{equation}
\label{potForm}
V(R) ~=~ \lim_{T\rightarrow \infty} {-1\over T} \ln {\int
d{\cal M}\exp (- S_W) \Pi
\over  \int d {\cal M} \exp (- S_W) }~. 
\end{equation}
where $\Pi$  
is the trace of an ordered product of $U$ matrices along
the perimeter of a long rectangle with sides of length $R$ and $T$, as
shown in Figure 4a.

\begin{figure}
\centerline{\psfig{figure=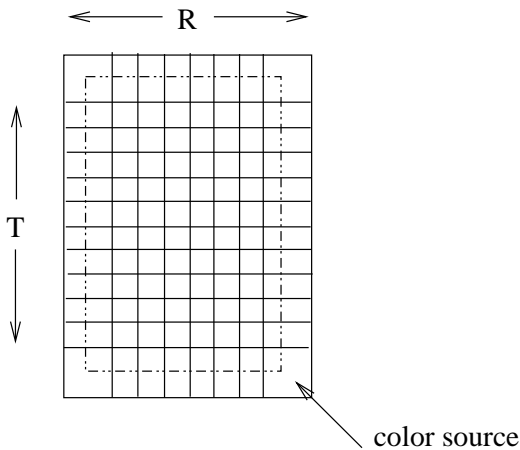,width=7cm}}
\caption[]{(a) The Wilson loop: insertion of a color (unit 
triality) source.}
\label{figfoura}
\end{figure}



With this background in hand, it becomes very easy to understand how
confinement arises in strong coupling.  The integral of a single $U$
matrix over the group manifold, $\int U^{-1}dU U$, vanishes.  (The
change of variables $U\rightarrow gU$ leaves the measure invariant, so
if $\int U^{-1} dU U = k$, then by changing variables
we find $gk = k$ for any $g$, which of course means $k=0$.)
So does $\int U^{-1} dU U^{-1}$.  Thus to find a non-zero contribution to
Eqn.~\ref{potForm} in strong coupling we must at least 
pull down plaquettes to share
sides with each of the links in the Wilson loop.

\begin{figure}
\setcounter{figure}{3}     
\centerline{\psfig{figure=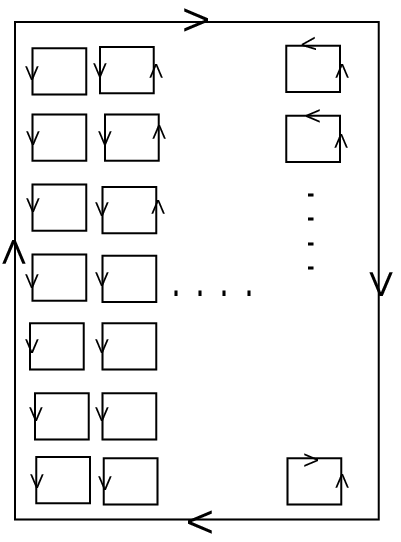,width=4cm}}
\caption{(b) Tiling of Wilson Loop in strong coupling.}
\label{figfourb}
\end{figure}



But now you see there's a whole new set of interior links with single
$U$ matrices and vanishing group integrals.  Clearly, to get a
non-zero contribution we must tile the whole area spanned by the
Wilson loop, as in Figure 4b.  This takes a number of plaquettes
proportional to the area $RT$, at least.  So the leading contribution
to the Wilson loop in strong coupling goes as
\begin{equation}
\label{potResult}
V(R) ~\sim~ - {1\over T}  \ln ({1\over g^2})^{RT} ~\propto R~.
\end{equation}
The linear potential, of course, means that it is impossible to
separate the color sources indefinitely, and so one has confinement.

\subsubsection{To the Continuum Limit}

The strong coupling result forms the starting point for a convincing
proof of confinement in (pure glue) QCD proper.  

One first argues that
the strong coupling perturbation theory has a finite radius of
convergence.  That can be done analytically.  Then one investigates
numerically whether there is a phase transition as a function of the
coupling, as the coupling varies from strong to weak.  It turns out
there
is a phase transition for $U(1)$, but not for 
$SU(3)$ (or $SU(2)$).  When there is no phase transition, the theory
remains in the same universality class, and its sharply defined
qualitative properties
cannot change. 

Thus in the physically relevant $SU(3)$ case:
\begin{itemize}
\item Since the strong coupling perturbation expansion converges, the
lowest non-trivial order governs the asymptotic behavior of the Wilson
loop, and exhibits confinement.

\item Since the strong coupling theory is in the  
same
universality class as the weak coupling theory, the weak coupling
theory also exhibits confinement.

\item Since asymptotic freedom implies that the weak coupling lattice
theory reproduces the continuum theory, the continuum theory exhibits
confinement. 
\end{itemize}

Now we see that is fortunate, and reassuring for this circle of ideas,
that there is a phase transition for
$U(1)$.  Otherwise we'd have proved confinement in QED, which would be
proving too much.

\subsubsection{Foundational Remarks}
To round out this discussion I would like to emphasize the deep
connections among renormalizability, asymptotic freedom, and lattice
gauge theory.  To construct a relativistic quantum theory, one
typically introduces at intermediate stages a cutoff, which spoils the
locality or relativistic invariance of the theory.  Then one attempts
to remove the cutoff, while adjusting the defining parameters, in
order to
achieve a finite, cutoff-independent limiting theory.  Renormalizable
theories are those for which this can be done, order by order in a
perturbation expansion around free field theory.  That formulation of
the problem of constructing a quantum field theory,
while convenient for mathematical analysis, obviously begs the
question whether this perturbation theory converges.  For interesting
quantum field theories, it rarely does.

A more straightforward procedure, conceptually, is to regulate the
theory as a whole by discretizing it.  This involves 
approximating space-time by a
lattice, and spoils the continuous space-time symmetries of the
theory.  Then one attempts to remove dependence on the discretization,
by refining it, while if necessary adjusting the defining parameters,
to achieve a finite limiting theory that does not depend on the
discretization, and therefore has a chance to 
respect the space-time symmetries.  The
redefinition of parameters is necessary, because in refining the
discretization one is introducing new degrees of freedom.  The
earlier, coarser theory results from integrating out these degrees of
freedom.  If it is to represent the same physics it must
incorporate their effects, for example in vacuum polarization.
Operationally, one can demand that some observable(s) measured at
scales well beyond the lattice spacing stays fixed as the
discretization is refined.  This fixes the free coupling(s).  The
question is then whether, having fixed the available parameters,  
the calculated values of {\it all\/} observables have finite limits.

This is very hard to prove, in general.  The only case in which it is
straightforward arises when
the effects of integrating out the additional
short-wavelength modes, that are introduced with each refinement of
the lattice, can be captured accurately by a re-definition of the
coupling parameter(s) already present in the theory.  That, in turn,
will occur in a straightforward way only if these modes are weakly
coupled.  For then perturbation theory will show us how to take the limit
for the renormalizable couplings, while assuring us that
naive power counting can be applied
to argue away all non-renormalizable ones.  But of course 
the ultraviolet modes
will be weakly coupled, if and only if the theory is asymptotically
free.  

Summarizing the argument, only those relativistic field theories which
are asymptotically free can be argued in a straightforward way to
exist.  Furthermore, the only asymptotically free theories in four space-time
dimensions involve nonabelian gauge symmetry, with highly restricted
matter content.  So the axioms of gauge symmetry and renormalizability
which we invoked to define QCD are, in a certain sense, redundant.
They are implicit in the mere {\it existence\/} of non-trivial
interacting quantum field theories.

\begin{figure}
\setcounter{figure}{3}     
\centerline{\psfig{figure=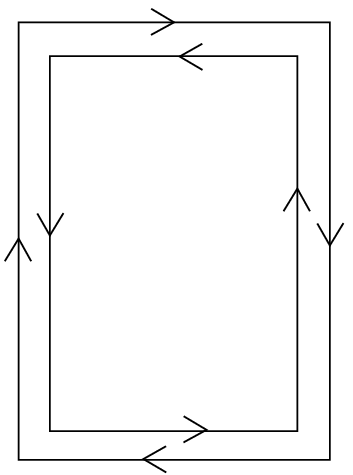,width=4cm}}
\caption{(c) Perimeter law, in the theory with quarks. }
\label{figfourc}
\end{figure}



\subsection{Chiral Symmetry Breaking}
\subsubsection{Numerical and Laboratory Phenomena}
The most direct evidence for chiral symmetry breaking in QCD comes
form numerical simulation of the theory.  One simply computes the
expectation value
\begin{equation}
\label{condensate}
\langle {\bar {q_L}}_i  {q_R^j}\rangle  \propto v \delta^j_i \neq 0
\end{equation}
in the ground state for the theory with massless quarks.  This
condensation, which breaks the chiral symmetry of the equations, is
entirely analogous to the development of spontaneous magnetization in
a ferromagnet.  As in that case, for any finite sample (e.g., in any
simulation) we must add an infinitesimal biasing field to stabilize a
particular alignment.

In Eqn.~\ref{condensate} I've chosen to align in the flavor diagonal
direction, but in the absence of a biasing field any chirally rotated
configuration, with $q_R \rightarrow Uq_R$, will have the same energy
but $\delta^i_j \rightarrow U^i_j$ in the expectation value, for any
$U \epsilon SU(f)$.  There are several technical issues in the
simulations that arise and must be addressed, but the numerical
evidence that chiral symmetry is spontaneously broken is unambiguous
and overwhelming, at least for $2\leq f \leq 4$.  I'll discuss this
evidence in more detail in Lecture 2.

The historical path whereby spontaneous chiral symmetry was discovered
as a property of Nature was of course quite different.  Indeed, the discovery
of chiral symmetry breaking in the strong interaction
antedates by more than a decade 
the discovery of QCD as its microscopic theory.

The conceptual starting point for the historic development was the
observation, coming into focus with the BCS theory of
superconductivity, that if a symmetry is spontaneously broken there
will be massless collective modes associated with this breakdown.
(The ``experimental'' starting point was the Goldberger-Treiman
relation; see below.)  Quite generally, suppose the ground state of a
physical system (e.g. a ferromagnet or the no-particle state of QCD at
zero temperature) is characterized by the existence of a condensate
\begin{equation}
\label{orderParam}
\langle \eta| {\cal M}^a_b| \eta \rangle = \eta^a_b
\end{equation}
that violates a continuous symmetry of the underlying equations
(e.g. rotational or chiral symmetry, respectively).  
Let the symmetry $g$ of the underlying
theory be implemented by the unitary operator $U(g)$, with $U(g)^{-1}
M U(g) = \rho(g) M$.   Then if $\rho (g) \eta \neq \eta$, which is the
signature of symmetry breaking, the states  
\begin{equation}
\label{degGrd}
U(g) |\eta \rangle ~\equiv~ | \rho(g) \eta \rangle
\end{equation}
will be physically distinct from the ground state, but energetically degenerate
with it.  By moving slowly within this manifold of states, as a function of
space, we would expect to create states whose energy goes to zero as
the wavelength of the variation goes to infinity.  In a particle
interpretation, the quanta of the field that creates such
configurations will be massless.  Furthermore, we have constructed a very
specific realization of these quanta in terms of symmetry generators.
This construction can be exploited to yield predictions for their properties.
For example, if the broken symmetry is an internal symmetry, the
quanta will be spin-0 particles.

Turning now specifically to the world of strong interactions, 
it is a striking fact that $\pi$ mesons are spin-0 particles which are much
lighter than all other hadrons.  This suggests the possibility
that they are associated with the spontaneous breakdown of an
approximate internal symmetry.  Their pseudoscalar character, and the fact that
they form an isotriplet, suggests the breaking pattern
\begin{equation}
\label{chiralPattern}
SU(2)_L \times SU(2)_R ~\rightarrow SU(2)_{L+R}~.
\end{equation}

This very specific picture of pions as collective modes closely
connected to broken chiral symmetry, besides explaining their quantum
numbers and small mass,  can be exploited to give many 
predictions about their low-energy behavior, as we shall discuss in
Lecture 2.  The phenomenological
success of these predictions validates
the hypothesis of spontaneously broken approximate chiral symmetry as
a description of Nature.  

Within QCD, this picture arises very naturally.
If the $u$ and $d$ masses are small the basic equations of QCD
will exhibit approximate chiral symmetry.  And numerical work 
QCD spontaneously
develops a symmetry-breaking condensate, as I mentioned.  
The general theoretical machinery
for extracting predictions from spontaneous symmetry breaking remains
valid and extremely valuable in modern QCD.  
Additionally, the specific form of
intrinsic breaking in QCD, through small quark mass terms, has specific
phenomenological consequences.  I will spell out how all this
works below in Lecture 2, when we discuss order parameters and
effective Lagrangians.  As we shall see, all these concepts
take on additional twists, and become even more central, for QCD in  
extreme conditions.

\subsubsection{Ironic Aside}
Ironically, the first generation of developments in high-energy
physics to be inspired by modern superconducitivity theory were inspired by 
BCS pairing
theory in the limit that the gauge coupling -- that is,
electromagnetism, and hence the phenomenon of superconductivity  --
is neglected.  In that limit it is the global
symmetry of electron number 
that is violated by the formation of a Cooper pair condensate, 
and there is a massless collective mode. Spontaneous breaking of a
global symmetry  turns
out to be the appropriate, and fruitful, idea for chiral
symmetry breaking in the strong interaction.

There was an interval of 
several years before the second generation of developments, when the
gauge coupling was reinstated.  Only then did the primary phenomenon of
superconductivity itself -- the Meissner effect -- enter the picture.
Rechristened in its new context as the Higgs mechanism, it 
of course became central to modern electroweak interaction theory.

\subsubsection{Pairing Heuristics}

Just as for confinement, the {\it fact\/} of spontaneous chiral
symmetry breaking in QCD is no longer negotiable.  Still, just as for
confinement, one would like
to understand why and how it occurs, and whether there are
circumstances in which it can come undone.  

A heuristic model for chiral symmetry breaking was supplied by Nambu and
Jona-Lasinio long before modern QCD.  Amazingly, with some re-labeling of the
players the concepts they introduced still apply.  Indeed, as we shall
see, at high density they come to look better than ever.  
We shall be discussing pairing
theory in great detail in Lecture 4, so this is just a foretaste.

Suppose one has an attractive four-fermion interaction 
\begin{equation}
\label{fourFermi}
{\cal L}_{\rm int.} ~=~ g (\bar \psi  \psi ) (\bar \psi \psi)~.
\end{equation}
Then one can imagine that it is energetically favorable to form a condensate 
\begin{equation}
\label{modelCond}
\langle \bar \psi \psi \rangle ~=~ v \neq 0~, 
\end{equation}
since this condensate generates negative interaction energy.  Indeed,
if the condensate is so large that we can ignore quantum fluctuations,
we shall have the condensation energy density
\begin{equation}
\Delta {\cal E} ~=~ - \Delta {\cal L} ~=~ 
- g \langle \bar \psi \psi \rangle^2 
\end{equation}

To test
this idea in the simplest crude way, write 
\begin{equation}
\label{crudeHubStrat}
(\bar \psi \psi )^2 ~=~ (\bar \psi \psi - v)^2 + 2 v \bar \psi \psi - v^2
\end{equation}
and, in the interaction Lagrangian, discard the fluctuating first
term (as is approximately valid at weak coupling).  The other terms, when 
added to
the standard kinetic energy
term for massless fermions, generate the Lagrangian for free 
massive fermions.  One can of course
diagonalize this quadratic approximate Lagrangian and, 
by filling the negative energy sea,
construct the appropriate zero fermion number density ground state (i.e., 
for a given
value of the condensate).  Then one can
enforce consistency by calculating $\langle \bar \psi \psi \rangle$ in
this state, and demanding that it is equal to the originally assumed value
$v$.  This consistency equation is
called the gap equation, in honor of its ancestor in BCS theory.  If
the gap equation 
has a non-trivial solution, one will have lowered the energy by
forming a condensate.

In QCD, the one-gluon exchange interaction is quite attractive in the
quark-antiquark color singlet channel.  This is hardly surprising,
since by forming a singlet one cancels the charge and eliminates field
energy.  To make a scalar condensate in this channel, one must pair
left-handed antiquarks with right-handed quarks.  So there are simple
heuristic reasons to anticipate the possibility of spontaneous chiral
symmetry breaking in QCD.

\begin{figure}[htb]
\centerline{\psfig{figure=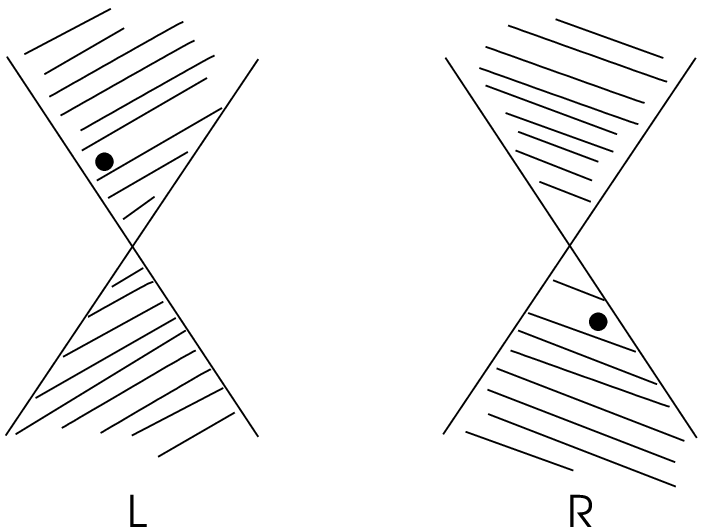,width=7cm}}
\vglue-.4in
\caption[]{Pairing picture for chiral symmetry breaking.  The dots
indicate the position of modes that pair.  In the limit of zero
energy, you must go 
the the tip of the light cone.}
\label{fig:5x.eps}
\end{figure}

Whether spontaneous chiral symmetry breaking actually occurs, however,
is a delicate question, because it involves a competition.  The
interaction energy one gains by pairing up occupied particle and
antiparticle modes must compete against the kinetic energy lost in
occupying them.  The kinetic energy can become arbitrarily small, but
only for a density of states that likewise vanishes -- the tip of the
Lorentz cone as shown in Figure 5.  So whether the interaction energy ever 
wins out, or
not, is a delicate dynamical issue.  For example, there is some
evidence that as the number of flavors $f$ grows the strength of
chiral condensation (relative the the primary QCD scale) shrinks, and
that for large enough $f$ 
(6? 7?)  it's gone.  This comes about,
presumably, because for larger $f$ the effective coupling does not run
so quickly to large values at small energy.

It's quite a different story at high density.  In that case the
density of states does not vanish, and spontaneous 
chiral symmetry breaking can appear at arbitrarily weak coupling,
where we have excellent analytic control.

\subsection{Chiral Anomalies and Instantons}
\subsubsection{The Historic Case}

The original discovery of the chiral anomaly involved what in
hindsight is a fairly complicated example of the phenomenon.  It came
about not through abstract investigation of mathematical models, 
but in the process of
analyzing a very specific physical process, the decay $\pi^0
\rightarrow \gamma \gamma$.
This is obviously an electromagnetic decay, so to treat it we must
consider QCD coupled to QED.  The combined theory, though it violates
isospin, still appears to be invariant under axial $I_3$ symmetry.
(I will  work, for simplicity, in the limit of vanishing $u$ and $d$
quark masses.  This can be shown to be a good approximation for
estimating the $\pi \gamma \gamma$ vertex, though of course the actual
$\pi^0$ mass must be inserted when we use this amplitude to 
calculate the decay rate.)  If this
were true, then the $\pi^0$ would still be accurately a collective
Nambu-Goldstone mode associated with the spontaneous breaking of axial
$I_3$ symmetry.  The coupling $\pi^0 F^{\mu \nu} {\tilde F}_{\mu \nu}$
would be forbidden, because at long wavelength such a Nambu-Goldstone
mode is derivatively coupled to the corresponding current.  Of course
one can rewrite $\pi^0 F^{\mu \nu} {\tilde F}_{\mu \nu} \rightarrow
{1\over 2} \partial^\mu \pi^0 A^\nu {\tilde F}_{\mu \nu}$ inside the
Lagrangian, after an integration by parts.  However the electromagnetic term
is not part of the axial $I_3$ symmetry current, classically.

The brilliant result of Adler, Bell and Jackiw is that when one
investigates the situation more deeply, using the full resources of
quantum field theory, such a term does in fact occur.  (Again, the
original analysis antedates QCD, and therefore was couched in rather
different language.  Its details are fascinating and of considerable
historical interest, but I will not discuss them here.)  Moreover its
coefficient can, given an underlying theory of the strong interaction,
be calculated precisely.  When the coefficient is calculated in a free
quark model -- with colored, fractionally charged quarks -- one finds
agreement between the predicted rate for $\pi^0 \rightarrow \gamma
\gamma $ and experiment.  Remarkably, the free-quark result remains
valid in QCD.

The mechanism whereby the extra term is generated is quite subtle.  It
is most readily seen in perturbation theory, though a non-perturbative
derivation is possible.

\begin{figure}[htb]
\centerline{\psfig{figure=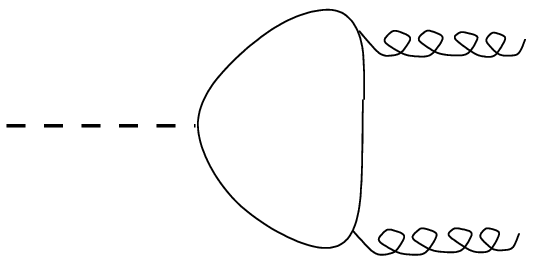,width=7cm}}
\vglue-.4in
\caption[]{Triangle graph, crucial for anomalies and Higgs
particle phenomenology.}
\label{fig:6fig.eps}
\end{figure}

To regulate the triangle graph and other loop graphs with circulating
fermions it is convenient to follow the procedure of Pauli and
Villars.  In this procedure, one introduces into the theory fictitious
boson fields $\phi$ carrying all the same quantum numbers as the
fermions (including spin ${1\over 2}$, and so flouting the
spin-statistics theorem) but having a large mass $M$.  Since the boson
loops differ in sign from the fermion loops their contributions will
cancel at very large virtual 
momentum, where the divergences arose.  Thus one
achieves finite integrals.  Then one takes the limit $M\rightarrow
\infty$, fixing a few low-energy amplitudes by setting them
equal to their
physical values.  
The basic result of
renormalization theory is that in suitable (renormalizable) quantum
field theories, having used a small number of low-energy amplitudes,
to determine the couplings -- which will be functions of the cutoff
and a few physically determined parameters --
all remaining physical amplitudes
will approach a finite limit, order by order in perturbation theory.  Of
course the limiting amplitudes no longer contain contributions arising
from the fictitious bosons as intermediate states, because those particles 
have been
driven to infinite mass.

This is not the place to review the technicalities of renormalization
theory.  Fortunately, for present purposes we don't need their details.  
Our focus is simply 
the triangle graph for the two-photon matrix element of the
divergence of the axial current.  
This graph, by naive power counting, is linearly convergent in the
ultraviolet by power counting.  In more detail, 
there are three fermion propagators, and gauge
invariance pulls out two powers of momentum to make $F$s from $A$s.
However, one will only be able to use gauge invariance if 
one has employed a gauge invariant
regulator, like Pauli-Villars. 

By the same power counting, the leading 
$M$ dependence of the regulated integral, which arises from the
Pauli-Villars boson loop, goes as
${1\over M}$ for large $M$.  This might seem to make it negligible.
The large mass $M$, however, involves a large violation of chiral
symmetry.  Specifically, it 
implies a large
coefficient $\propto M$ in the divergence of the axial
current:
\begin{equation}
\label{bigMDiv}
\partial_\mu (\bar \phi \gamma^\mu \gamma^5 \phi ) ~=~ 2M \bar \phi
\gamma^5 \phi~. 
\end{equation}
(To have a regulated chiral symmetry that will behave sensibly as we
remove the cutoff, the Pauli-Villars fields must transform in the same
way as the quark fields they regulate.)  This factor exactly cancels
the convergence factor, leaving a finite residual contribution.  The
regulator, necessary to control the contribution of highly virtual
quarks, spoils naive chiral symmetry.  That is the essential mechanism
of the chiral anomaly.

\subsubsection{An Easier Version}
 
The basic mechanism leading to anomalies is nicely illustrated, in a
context where it is relatively easy to understand, by the process $h
\rightarrow GG$ of Higgs particle decay into two gluons.  Since this
process is of independent and timely interest, I hope you'll forgive a
brief diversion.

The coupling of the standard model Higgs boson to quarks is given as
\begin{equation}
\label{hCoupling}
{\cal L}_{\rm int}~=~ -2^{1\over 4}G_F^{1\over 2}  h\sum m_i {\bar q}_i q_i~,
\end{equation}
where $G_F$ is the Fermi coupling constant.
The direct coupling to ordinary matter would seem to be extremely
feeble, due to the very small masses of the $u$ and $d$ quarks.
Ordinarily one would expect that the contribution from the heavy
quarks would be suppressed, according to general ``decoupling''
theorems.  Indeed, we'd be in pretty bad shape in physics if we
always had to worry about big contributions to low-energy amplitudes
from (potentially unknown) heavy particles.  However there is an
exception of sorts here, because the coupling grows with the mass.
Thus the contribution to the dimension 5 induced interaction term
\begin{equation}
\label{inducedL}
{\cal L}_{\rm induced} ~=~ \kappa h ~{\rm tr}~ G^{\mu \nu} G_{\mu \nu}
\end{equation}
arising through a heavy quark loop of circulating heavy quarks in the triangle
graph of Figure 6, which by power counting one might expect to be
inversely proportional to the mass of the quark, is instead
unsuppressed.  (Of course, this time the external legs are a Higgs
particle and two gluons.)  This is very similar to what we saw for the
triangle anomaly, but now with real heavy particles rather than
virtual regulators.  One finds
\begin{equation}
\label{inducedK}
\kappa ~=~ 2^{1\over 4} G_F^{1\over 2} {g^2\over 48 \pi^2}
\end{equation}
per heavy quark, in the large mass limit ($m_h \lsim m_q$).  For
values $90~ {\rm Gev }\lsim m_h \lsim 150~ {\rm Gev}$ of the Higgs mass
most interesting for ongoing searches this ``anomalous'' 
gluon coupling generates both the
dominant mechanism for hadronic production of $h$ particles, 
and a significant decay
mode for them.

\subsubsection{The Saga of Axial Baryon Number}

For concreteness in this part I'll mostly take $f=2$ and refer to
the quarks as $u$ and $d$.  This simplifies the notation, loses
nothing essential, and is close to reality.

The spontaneous breakdown of approximate chiral $SU(2)_L\times SU(2)_R
\rightarrow SU(2)_{L+R}$ is associated with an extensive, successful
phenomenology.  At the level of quarks in QCD, we understand it as the
result of the development of a condensate $\langle \bar {u_L} u_R
\rangle ~=~ \langle \bar {d_L} d_R \rangle$.  This condensate also
violates the axial baryon number symmetry under $(u_L, d_L)
\rightarrow e^{i\alpha} (u_L, d_L)$, $(u_R, d_R) \rightarrow
e^{-i\alpha} (u_R, d_R)$, which is a symmetry of Eqn.~\ref{masslessL}. 
One might expect, then, a light Nambu-Goldstone boson with the
associated properties -- a light, flavor-singlet pseudoscalar with
highly constrained couplings.  Alas, there is no such particle.
Thereby hangs a tale.

The first observation is that there is an ``anomalous''
contribution to the divergence of the axial baryon number current,
arising from the triangle graph of Figure 6.  There's a new set of
players -- the axial baryon current instead of axial $I_3$, and gluons
instead of photons -- but they follow the same script.  Thus we find
\begin{equation}
\label{aFiveAnom}
\partial_\mu j_B^{\mu 5} ~\equiv~ \partial_\mu (\bar  u \gamma^\mu
\gamma^5 u + \bar d \gamma^\mu \gamma^5 d) ~=~ {g^2\over 4 \pi^2} {\rm
tr}~ {\tilde G}_{\mu\nu}G^{\mu\nu}~,    
\end{equation}
where 
${\tilde G}_{\mu\nu}\equiv {1\over 2} 
\epsilon_{\mu\nu\alpha\beta}G^{\alpha\beta}$ is the dual field strength.
The current $j_B^{\mu 5}$ is not conserved.  Naive axial baryon
number symmetry, generated by the spatial integral of $j_B^{0 5}$, is
spoiled by an anomaly.

However, the existence of this anomaly does not in itself remove the
problematic mode.  For one has
\begin{equation}
\label{divK}
{\rm tr}~G^{\mu \nu} {\tilde G}_{\mu \nu} ~=~ \partial_\mu K^\mu
\end{equation}
where 
\begin{equation}
\label{kDef}
K_\mu ~=~ {1\over 2} \epsilon_{\mu\alpha\beta\gamma}
{\rm tr}~
(A^\alpha\partial^\beta A^\gamma + {2\over 3}A^\alpha A^\beta A^\gamma) ~.
\end{equation}
Thus it would appear that a modified symmetry, generated by the charge
associated with the modified conserved current $j^{A5}_\mu - K_\mu$,
is spontaneously broken, leaving us not much better off than before,
still with the mistaken prediction of an extra light flavor singlet
pseudoscalar.

Fortunately, though, Eqn.~\ref{divK} is itself problematic.  The point is
that $K_{\mu}$ is a gauge-dependent quantity, so that in principle it
can be singular without implying the singularity of any physical
observable.

We can be more specific about this in the context of a path integral
treatment of the theory.  In such a treatment we express quantum
amplitudes as an integral of contributions from different classical
field configurations.  To have reasonable control of the functional
integrals -- to have a measure that is damped for large field
strengths -- we must consider the Euclidian form of the theory,
rotating to imaginary values of the time.  In this framework, consider
the behavior of the gauge potentials $A_\mu$ at spatial infinity.  For
the integral of $K_\mu$ to acquire a non-vanishing surface term, and
thereby violate the formal conservation law, requires that $A_\mu(x)
\sim {M_\mu(\hat x) \over |x|}$.  This is not allowed for generic
forms of $M_\mu(\hat x)$, since a non-trivial field strength $G \sim
{1\over r^2}$ would result, leading to a logarithmically divergent
action.  But for special values of the boundary conditions $M_\mu(\hat
x)$ these can cancel, leaving behind a finite-action contribution to
the functional integral that contributes to $\partial_\mu K^\mu$.

In the weak-coupling limit, we look for the field configurations that
contribute to the amplitude of interest that have  the
smallest possible action.  
For amplitudes that violate axial baryon number (non-vanishing $\int
G^{\mu \nu} \tilde G_{\mu \nu}$) these configurations
are instantons, much-studied
classical solutions of the Euclidian Yang-Mills equations.

A very important consistency check is that the integral $\int {\rm
tr}~G^{\mu \nu} \tilde G_{\mu \nu}$, which according to the anomaly
equation measure the violation of axial baryon number, must turn out
to be quantized (that is, come in discrete units -- it's a c-number
integral).  Indeed, since axial baryon number is quantized, 
changes in axial baryon number had also better be quantized.  
The facts we have discussed,  that this integral can be thrown onto the
surface at infinity and that its finiteness requires special conspiracies,
suggest a connection to topology and the possibility of its quantization.
These features can indeed be demonstrated, but I won't do that here.  

In
view of the easily proved inequality
\begin{equation}
\label{topoAction}
\int {\rm tr}~G^{\mu \nu} G_{\mu \nu} \geq | \int {\rm tr}~G^{\mu \nu}
\tilde G_{\mu \nu} | 
\end{equation}
quantization of the right-hand side implies that
the action of any configuration leading to axial baryon
number violation is bounded below by a finite constant.  Thus in
the function integral such configurations 
are suppressed by a factor $e^{-{1\over g^2} {\rm
const.}}$.  In particular, they vanish to all orders in perturbation theory!

Instantons violate axial baryon number, but none of the other chiral
flavor symmetries.  This allows us to visualize their effect quite
simply, at a heuristic level.  The flavor structure must contain a
product of determinant factors
\begin{equation}
\label{instFlavor}
{\cal L}_{\rm inst.} ~\propto~ \epsilon_{ij} q_L^i q_L^j  \epsilon^{kl}
\bar{q_R}_k \bar{q_R}_l~. 
\end{equation}
For $f$ flavors, we'd have $f$ lefties in and $f$ righties out.  The
structure of the color and spin indices, and the form-factor
accompanying this interaction, are more tricky to work out.  To do it
one must solve for the fermion zero-modes in the presence of an
instanton.  The resulting effective non-local ```tHooft
interaction'' is represented pictorially in Figure 7:  lots of
fermions emerging from an extended gluon cloud.

\begin{figure}[htb]
\centerline{\psfig{figure=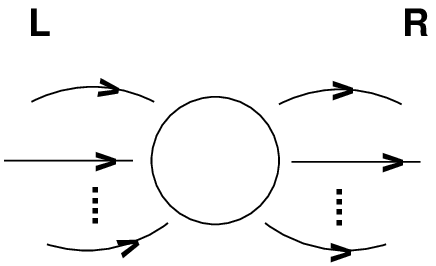,width=7cm}}
\vglue-.4in
\caption[]{`tHooft interaction. Chirality is violated by a knot of
gluon energy.}
\label{fig:seven.eps}
\end{figure}

Unfortunately, for reasons we have discussed before, 
the formal weak-coupling limit that leads us to focus
on instantons is not under
control for QCD in vacuum.  So the arguments used in this section, while
certainly suggestive, cannot be made quantitative in a
convincing way.  (Nevertheless, some reasonably successful
phenomenology has been done by the saturation of the functional
integral by superpositions of instantons and anti-instantons as a
starting point.)    
Remarkably, under extreme
conditions the theoretical situation for axial $U_A(1)$ breaking
comes under much better control,
with interesting consequences, as we'll see.

\section{Lecture 2:  High Temperature QCD: Asymptotic Properties}

\subsection{Significance of High Temperature QCD}

In this lecture I shall be discussing the behavior of QCD (and some
closely related models) at high temperature and zero baryon number
density.  Since that is a mouthful I'll just say high temperature.

The high temperature phase of 
QCD is of interest from many points of view.  
First of all, it is
the answer to a fundamental question of obvious intrinsic interest:
What happens to empty space, if you keep adding heat?  Moreover, the
high density phase of QCD 
was (almost certainly) the dominant form of matter during the
earliest moments of the Big Bang.  Moreover, it is a state of matter
one can hope to
approximate, and study systematically, in heavy ion collisions.
Major efforts are being directed toward this goal and significant,
encouraging 
results have already emerged.  One can also simulate many aspects of
the behavior
with great flexibility and control, from first principles, 
using the techniques of lattice
gauge theory.  One can also make progress
analytically.  So there is a nice interplay among physical experiments,
numerical experiments, and theory.     

The fundamental theoretical result regarding the asymptotic high
temperature phase is that it becomes quasi-free.  That is, one can
describe major features of this phase quantitatively by modeling it as
a plasma of weakly interacting quarks and gluons.  In this sense the
fundamental degrees of freedom of the microscopic Lagrangian,
ordinarily only indirectly and very fleetingly visible, become
manifest (or at least, somewhat less fleetingly visible).  Likewise
the naive symmetry of the classical theory which, as we saw in Lecture
1, is vastly reduced in the familiar, low-temperature hadronic phase,
gets restored asymptotically.  In particular, chiral symmetry is
restored, and confinement comes completely undone.  Axial baryon
number and scale symmetry, though never precisely restored, become
increasingly accurate.

Since there are dramatic qualitative differences between the
zero-temperature and the high-temperature phases, the question
naturally arises whether there are sharp phase transitions separating
them, and if so what is their nature.  This turns out to be a rich and
intricate story, whose answer depends in detail on the number of
colors and light flavors.  In the course of addressing it, we shall
have to refine and modify common, rough intuitions about chiral
symmetry and (especially) confinement.   After an involved but I think
interesting and coherent story, building up from the study of 
various idealizations,
we shall find that there is,
plausibly, a true phase transition in real QCD that we can 
converge upon from several directions --  
experimentally, numerically, and analytically.

\subsection{Numerical Indications for Quasi-Free Behavior}

For technical reasons it has been difficult until recently 
to simulate QCD including
dynamical quarks with realistically small masses.  That situation is
changing, but it will be a few years before accurate quantitative results for
thermodynamic quantities for QCD with light dynamical quarks become 
available.  

\begin{figure}[!t]
\centerline{\psfig{figure=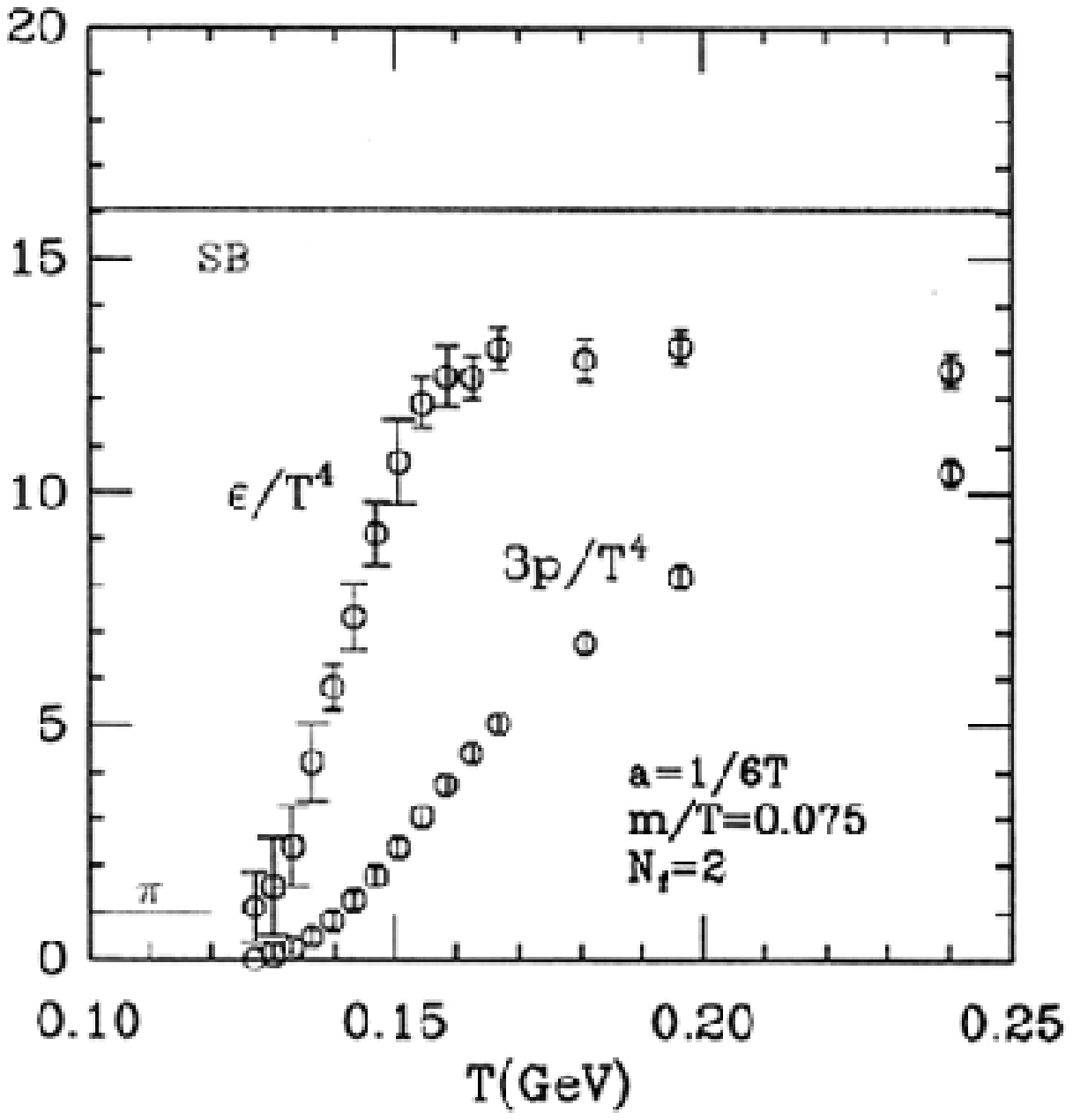,width=7cm}}
\vglue-.4in
\caption[]{Energy density and pressure of 2-flavor QCD as a function
of temperature. } 
\label{fig:working-top.ps}
\end{figure}

Fortunately, we can already learn a lot from the existing simulations
using pure glue or glue plus moderately massive quarks.  

Representative results for the temperature dependence of the energy
density and pressure in the two flavor theory are shown in Figure 8.
Clearly, there is a rapid crossover in the behavior, with dramatic
rises in the energy density and pressure (even when normalized to
$T^4$) over a small range of temperatures around 150 Mev.

A notable feature of the numerical results is that while the energy
density (divided by $T^4$) ascends rapidly to something close to its
asymptotic value, the pressure appears much more sluggish.  Thus the
behavior of the plasma, even in regard to this basic bulk property,
differs significantly from a free gas of massless particles.  It is a
worthy challenge to compute the corrections to free behavior
analytically in weak coupling.  This is not entirely straightforward,
due to the absence of magnetic screening in perturbation theory
(concerning which, more below).  For recent progress see \cite{ABS}.

In reality the only hadrons light on the scale of the computed
cross-over temperature are pions.  Thus for temperatures significantly
below this temperature (say $T \leq 120~ {\rm Mev}$) one has a rather
dilute gas of pions, with 3 massive degrees of freedom for the
three possible charge states of these spinless particles.
Asymptotically, on the other hand, one has a gas with three different
flavors of quarks, each of which comes with two spins, three colors,
and antiquarks.  Also there are eight gluons, each with two
helicities.  Thus the number of degrees of freedom is $3\times 2\times
3\times 2 + 8\times 2 = 52$, of which all but the strange quarks are
essentially massless.   Evidently, the difference is gigantic!
Remarkably, the change from one regime to the other 
appears to occur largely within a narrow range of 
temperatures around 150 Mev, amazingly low if regarded from the
hadron side.

\subsection{Ideas About Quark-Gluon Plasma}

The physics of quark-gluon plasma is already a big subject with a vast
literature.  It will bloom further as the RHIC and ALICE programs
gather data.  Let me briefly sketch a few of the characteristic
phenomena that
have been discussed.
 
\begin{itemize}
\item A fundamental foundational result is the observation that
distributions of particle energies in the final state are well
described by thermal distributions corresponding to a freeze-out
temperature around 120 Mev.  This observation makes it extremely
plausible, as had already been anticipated from theoretical work, that
approximate thermal equilibrium (at least, kinetic equilibrium) is
established -- at higher temperatures, of course -- in the initial
fireball.  That's very good news, both because it makes the
theoretical analysis easier, and because it means that the collisions
really are approximating the conditions which are of most fundamental
interest.

\item The most basic and profound prediction is what I have already
mentioned, that one should have approximately the energy and pressure
characteristic of the appropriate -- large! -- number of microscopic
degrees of freedom.  Qualitatively, this means among other things a
steep rise in the specific heat, so that the rise of temperature with
energy will slow markedly.  Energy will go into particle production,
not motion. In principle, the temperature is
accessible either through measurement of the transverse momenta of
hard leptons or photons emerging from the initial fireball.  The
entropy can be estimated from the final thermal particle distribution
at freezeout, since the expansion and cooling should be roughly
adiabatic until freezeout.  Several more sophisticated flow
diagnostics have been proposed to give handles on the full equation of
state.

\item Since strange quarks are expected to be much lighter than the lightest
hadrons ($K$ mesons) in which they are found, one can anticipate a
significant rise in the relative 
multiplicities of strange (and antistrange)
particles, relative to normal hadronic collisions.   There is
already a striking phenomenon of this kind seen at SPS, with a
dramatic rise (as much as a factor 15) 
in $\Omega$ and $\bar \Omega$ production.

\item Perhaps the single most striking experimental result to emerge so far
from the study of heavy ion collisions is the suppression of J/$\psi$,
relative to Drell-Yan background, in lead-lead collisions at the
highest energies.  When the ratio is plotted as a
function of atomic weight and energy, a clear break in its behavior,
relative to lower energies or lower atomic numbers, leaps
to the eye.  No hadron-based model of the collision process
anticipated this break, and
none has been successful in reproducing it.  On the other hand an
effect of just this sort was anticipated, based on simple qualitative
arguments, to mark the onset of quark-gluon plasma behavior.  

The basic point is that free gluons are very effective in dissociating
J/$\psi$ particles.  In quark-gluon plasma there is an abundance of
free gluons, while in the hadron phase there is a large mass gap for
glue.  Alternatively, we may say that in the plasma phase color
screening prevents J/$\psi$ binding.  Unfortunately, while some effect
of this kind is very plausible, and evidently does occur, it seems
difficult to refine the heuristic argument into a really precise
calculation.

\item The best hope for a rigorous characterization of quark-gluon
plasma behavior is probably comparison of experimental measurements to
calculated predictions for quantities that can be addressed using
(sophisticated extensions of) perturbative QCD.  Among the most
promising candidates are hard probes, such as high transverse momentum
jets, high-mass dileptons, and energetic photons.  Assuming thermal
equilibrium -- or a definite model of quasi-equilibrium -- one can
formulate reasonably precise expectations for the rates and
distributions of these phenomena, with many cross-checks.  Their use
is analogous to the use of radiative probes in traditional plasma
diagnostics.  Another characteristic 
signature of quark-gluon plasma is softening of quark
jet distributions due to their passage through the medium.
\end{itemize}

A more conjectural possibility, which has received much attention under
the name ``disordered chiral condensate'' or DCC, is
that the return to equilibrium as the fireball cools is marked by
collective relaxation.  Then one might see 
gross deviations from equipartition in the
collective modes.

Specifically, as we shall discuss at length below, one expects that
the large spontaneous breaking of chiral
symmetry which occurs in the ground state comes undone at high
temperatures.  The transition from chiral symmetry breaking to chiral
symmetry restoration is described by equations very similar to the
equations that describe the loss of magnetization when one 
heats a magnet past its Curie
temperature. 

To make the analogy accurate 
we must envision the demagnetization taking place
in the presence of a tiny external field, representing the small
intrinsic breaking of chiral symmetry due to non-zero $u$ and $d$
quark masses.  Now if, after being disordered at high temperature,  
the magnet is cooled rapidly there are two
extreme possibilities for how it might relax back down to the ground state.
According to  one extreme picture, 
each spin separately and independently settles
down to align with the external field.  According to  
the other extreme picture,
the spins first align with each other in large clumps, 
usually in some
wrong direction, before the clumps relax collectively, as units, toward the
correct alignment.  In the latter case, one will have significant 
correlation phenomena, and the possibility of 
coherent radiation.  In QCD, this will take the form of ``pion lasing'' --
an abnormally large number of pions occupying a small region of phase
space, with a highly non-Gaussian distribution of charged to neutral
multiplicity, will emerge.

Because it is an intrinsically non-equilibrium phenomenon, the
likelihood of DCC formation is hard to assess theoretically.  It
has been observed in some idealized numerical simulations.

Evidently there is a plenitude of signature phenomena in quark-gluon
plasma that can and presumably will be explored in heavy ion
collisions.  We can look forward to a many-faceted dialogue between
theory and experiment in coming years.

For the remainder of this lecture and in the next one, however, I will
focus on the specific, narrower theoretical question of equilibrium
phase transitions.  This emphasis brings the advantage that the
conceptual issues become well-posed and precise, and support a rich
theory; on the other hand  application of the results derived to the
complex realities of heavy ion collisions is not straightforward.

\subsection{Screening Versus Confinement}

One should not assume, that because the quark-gluon plasma at high
temperatures is
conveniently described using very
different degrees of freedom from those we use to describe
the hadronic gas at low temperatures,
there must be a sharp phase transition separating them.  Indeed, ordinary
plasmas are very different from gases of atoms (so different, that
at Princeton they are studied on different campuses), but it is well
understood that no strict
phase transition separates them.  The fraction of ionized atoms rises
smoothly, though rather abruptly, 
from nearly (but not quite) zero at low temperatures to nearly unity
at high temperatures.

With this cautionary example in mind, let us revisit the question of
confinement.  Previously we discussed the pure glue theory, and were
able to give a precise definition of confinement in terms of the
asymptotic behavior of Wilson loops.  We were even able to understand
in a very simple way why confinement is not at all a bizarre or
mysterious behavior, but quite a reasonable possibility for a
strong-coupling (or asymptotically free) gauge theory.  Actually, it
was lack of confinement that required some explaining away--a failure of
the strong-coupling expansion, or a phase transition.

Now let's consider the theory with quarks.

The strong coupling expansion requires that we use the discretized
lattice version of the theory.  The basic idea of its
extension to include quarks is
quite simple, although there are great subtleties if one tries to
do justice to chiral symmetries, and many algorithmic issues. 
These questions involve important,
active areas of research.  However they do not impact the basic issues
of screening versus confinement, as discussed in this section.
  
To give the quarks dynamics, we need to supply a `hopping' term.
The sum over all links of 
\begin{equation}
\Delta {\cal L}_{\rm hop} ~=~ 
\psi(\hat r) U_{r,\mu} \bar \psi(\hat r +\mu)
\end{equation}
does the job, and reduces formally to the continuum action for $a$
small.  It has an evident gauge invariance, generalizing
Eqn.~\ref{transformU}, whereby the $\psi$ variables, which live on
vertices, are simply multiplied by the corresponding $\Omega$s.

Revisiting the question of tiling the Wilson loop, we see that now it
is possible to get a non-zero contribution by propagating a single
quark line around the perimeter, as shown in Figure 4c.  This is quite
unlike the
pure glue theory, where we were required to tile a whole area.  The
perimeter tiling
corresponds to a potential which does not continue to grow at large
distances, but rather
saturates at a
finite value.  Physically, it corresponds to the production of a
separated meson pair.  The color sources, inserted by the two sides of
the Wilson loop, can be saturated by a dynamical quark on one side, and a
dynamical antiquark on the other.   There is a finite energy to make
the pair, but once it is 
made and combined with the sources into `mesons',
the mesons have only short-range residual
interactions, and the total energy  does not grow with the distance.

There is a simple heuristic way to understand the difference between
the two cases.  There is an additive quantum number modulo 3,
triality, characterizing color
charges.  It is one for quarks, minus one for antiquarks, and zero for
gluons.  If we write $SU(3)$ indices on the fields, triality is simply
the number of upper indices minus the number of lower ones.  Because
of the existence of the invariant epsilon symbol $\epsilon^{abc}$ 
triality can jump in units of three by color invariant processes, but
not in units of one or two.   
In the pure glue theory all the dynamical fields
have zero triality, so a source of unit triality cannot be screened.
Furthermore the 
presence, or not, of unit triality 
can be determined by measurements made at great distances.
We saw this in the strong coupling expansion.  A triality source 
generated a `live'
link that could be displaced by laying down plaquettes, 
but not cancelled.  We have, therefore, a poor man's version of Gauss'
law.  If triality flux interferes with the correlations in
the ground state, then
as we
separate source and antisource we will produce 
a finite change in vacuum energy per unit volume that extends over a
growing volume, with confinement a conceivable outcome.  By contrast, in
the theory with dynamical quarks triality can be screened.  In the
absence of any strictly conserved quantity characterizing a source, it is
difficult to imagine how its dynamical influence could extend to great
distances.  In fact, it would be hard to specify exactly what it is
that is confined.

By the way, if the only dynamical quarks are extremely heavy ones then
the area tiling can remain cheaper than the perimeter tiling up until
very large values of the separation $R$.  In this case one will have a linear
interquark potential out to large $R$, supporting a spectrum of bound
states up to an ionization threshold.

\subsection{Models of Chiral Symmetry Breaking}

To help ground our later discussions, I will now briefly discuss some
basic elements of the phenomenology of chiral symmetry breaking in
the observed strong interaction, and in QCD.

The circle of ideas around chiral symmetry breaking grew up around
attempts to understand a remarkable formula discovered by Goldberger
and Treiman.  Their derivation of the formula made use of drastic and
uncontrolled approximations, and is mainly of historical interest.
The modern understanding starts from ideas introduced by Nambu and
Gell Mann and Levy, and developed with great ingenuity by many
physicists.  Their hypotheses are fully justified within QCD.  Indeed,
nowadays it is appropriate to start from QCD, and to interpret
the necessary hypotheses within the microscopic theory.

Interpreted within QCD, the hypothesis of chiral symmetry
breaking has two parts:

i. The $u$ and $d$ quark masses are small, so that the corresponding
fundamental interaction terms $m_u \bar u u$ and $m_d \bar d d$ in the
Lagrangian may be treated as perturbations.

Thus we are invited to consider the properties of a zeroth-order
theory with massless $u$ and $d$ quarks.  In this limit, as we have
discussed, there is an $SU(2)_L \times SU(2)_R$ chiral symmetry of the
fundamental theory, rotating among the different helicities
separately.

ii. In the absence of $u$ and $d$ quark masses, the $SU(2)_L \times
SU(2)_R$ chiral symmetry is spontaneously broken, down to the diagonal
vector subgroup $SU(2)_{L+R}$.

More precisely, the hypothesis is that a condensate 
\begin{equation}
\langle \bar u u \rangle = \langle \bar d d \rangle = v \neq 0
\end{equation}
develops. 

One can also consider extending these hypotheses to the $s$ quark, but
it is not entirely clear under what circumstances it is safe to treat
$m_s \bar s s$ as a perturbation.

A consequence of these hypotheses is that one expects the existence of
approximate Nambu-Goldstone bosons.  If it were an exact symmetry that
were spontaneously broken we would have exactly massless particles of
this type; since there is some small intrinsic breaking, in addition
to the larger spontaneous breaking, the corresponding Nambu-Goldstone
particles acquire non-zero, but small, masses.

There are indeed particles within the observed hadron spectrum that
are much lighter than any of their brethren, namely the $\pi$ mesons.
Furthermore the quantum numbers of the $\pi$ mesons -- $J^{PC} =
0^{-+}$, $SU(2)_{L+R}$ (isospin) triplet -- are what one requires for
Nambu-Goldstone bosons arising from $SU(2)_L \times SU(2)_R
\rightarrow SU(2)_{L+R}$ breaking.

To see this, consider the physical origin of the Nambu-Goldstone
bosons.  They arise due to the possibility of obtaining low-energy
field configurations by interpolating slowly, in space and time, among
the energetically degenerate but inequivalent ground states one has
due to spontaneous symmetry breaking.  The inequivalent ground states
are generated by three independent transformations of the type $(q,
-q)$ in the Lie algebra of $SU(2)_L\times SU(2)_R$, which manifestly
form an isotriplet of odd parity.  Furthermore there is no preferred
space-time direction in the condensate, so the quanta are spin 0.

Although it is fundamentally a phenomenon of the strong interaction,
much of the interest of chiral symmetry derives from its connection
with the weak interaction.  Specifically, the currents that generate
the approximate chiral symmetry of the strong interaction also appear
in the weak interaction.  The prototype application, the
Goldberger-Treiman relation, exploits this connection.  The pion decay
$\pi^+ \rightarrow \mu^+\nu$ involves the hadronic matrix element of
the axial vector current 
\begin{equation}
\langle 0 | A^5_\eta | \pi^+
\rangle = F_\pi p_\eta 
\end{equation} 
where $p$ is the momentum.

Thus $F_\pi$ is a directly measurable quantity.  For the divergence of
the axial current we find then 
\begin{equation}
\label{axi} 
\langle 0 |\partial^\eta A^5_\eta | \pi^+ \rangle~=~ F_\pi m_\pi^2~.
\end{equation} 
We see here the connection between chiral symmetry and
the mass of the pion: in the version of QCD with exact chiral symmetry
the divergence would vanish, and so would the mass of the pion.

Now let us consider another matrix element that appears in describing
another basic  weak
process, that is beta decay of the neutron.  The nucleon matrix element of the
axial current 
\begin{equation} 
\langle N | A^5_\eta | N \rangle ~=~
G_A \bar u \gamma_\eta \gamma_5 u 
\end{equation} 
at small momentum
transfer for nucleons nearly at rest.  $G_A$ is a quantity subject to
strong-interaction corrections and it is therefore 
not the sort of thing we can normally
expect, in the absence of special insight, to calculate easily.  It is
measured to be about 1.2.  Taking again the divergence, we have on the
right-hand side $2M G_A\bar u \gamma_5 u$, not particularly small
(beyond the kinematic suppression), whereas on the right-hand side we
have the matrix element of a ``small'', chiral-symmetry breaking
operator.  However there is a contribution to this matrix element arising
from the nucleon coupling to a $\pi$ meson, which then communicates with
the current divergence according to Eqn.~\ref{axi}.  The Feynman graph for 
this is
shown in Figure 9.  The factors of $m_\pi$ cancel, and we find

\begin{equation}
\label{gtr} 
g_Y F_\pi ~=~ 2 m_N G_A~, 
\end{equation} 
which is the
Goldberger-Treiman relation.

\begin{figure}[htb]
\centerline{\psfig{figure=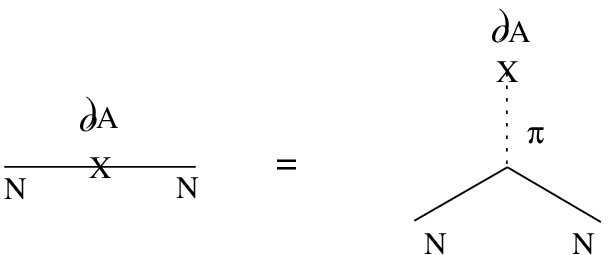,width=10cm}}
\vglue-.4in
\caption[]{ Saturation of the axial current divergence with the nearby
$\pi$.}
\label{fig:ninefig2.eps}
\end{figure}

This logic of this derivation of the Goldberger-Treiman relation can
be vastly generalized, to include matrix elements of 
low-momentum pions or currents between various states.
It can be made systematic by using the
technology of Ward identities.  In this context one finds that in
relating multi-current Green functions to multi-pion processes one
must often evaluate current commutators.  One obtains in this way 
a host of predictions
for low-energy processes, which work remarkably well.  (At high
energies or momenta the saturation of axial currents with pions is no
longer accurate.)  The successful evaluation of
such commutators, in agreement with experiment, 
using relations abstracted from free field theory
(now justified in QCD), was a major step in the historical elucidation
of the strong interaction, and in the revival of interest in quantum
field theory in the late 1960s.  When I reflect that this elaborate 
theoretical
technology was developed before the microscopic theory, by working
backward from nuggets of relevant data embedded within an
overwhelming confusion of strong-interaction phenomena, I am lost in
admiration.

Many of the results of Ward identity and current algebra gymnastics
can be derived in an easier, more transparent way by writing down appropriate
effective Lagrangians.  If they embody the correct broken and unbroken
symmetries, such Lagrangians will satisfy all the Ward identities that
can be derived as consequences of these symmetries.  Thus one can
reproduce more transparently 
the valid results of the Green function analysis -- and
more.  Generally the `more' -- relations derived from the effective
Lagrangian away from the low-energy
limit -- will depend on non-generic features of the
effective Lagrangian, and should be ignored.

\subsection{More Refined Numerical Experiments}

With this background, we are now in a better position to discuss two
more refined diagnostics of finite temperature QCD.  

The first is the so-called Polyakov loop.  It is basically half a
Wilson loop.  Let me be more precise. 
A standard result in path integral theory states that one
can set up the partition function at finite temperature by passing to
imaginary values of the time variable and requiring periodicity of the
fields (antiperiodicity for fermions) under $\tau \rightarrow \tau +
1/T$.   Now we can consider parallel transport around the circle of
imaginary time:
\begin{equation}
\langle L  \rangle ~ =~ \langle~
{\rm Tr}~\exp\{ i\int_0^\beta A^a_\tau \lambda^a d\tau \}~ \rangle  .
\end{equation}

By the same strong coupling argument as before, we anticipate that the
expectation value of the Polyakov loop integral will vanish
in a confining phase.

It is very pleasant that we can (following
Polyakov) relate this anticipation to a symmetry principle.  The action
for the pure glue theory is left invariant if we multiply all the $U$
matrices connecting (say) $\tau = 0 $ to their temporal neighbor 
$\tau = a$ vertices by an
element of the center of the group, since any plaquette product has
either no such $U$ or one going up and one down.  Thus for $SU(2)$ we
can multiply by -1 (times the identity matrix), 
and we have the symmetry $Z_2$, whereas for $SU(3)$ we can multiply by
$\omega$ or $\omega^2$, where $\omega = e^{2\pi i/3}$ is the cube root
of unity, and we have the group $Z_3$.  On the other hand the Polyakov
loop is {\it not\/} left invariant under these operations, but rather
multiplied by the corresponding numerical factor.  If the expectation
value of the loop vanishes, as is characteristic of the confined
phase, then the confinement symmetry is valid.  
However if the expectation value of
the Polyakov loop does not vanish the confinement symmetry is spontaneously
broken.   

The discrete ``confinement symmetry'' of the pure glue theory is {\it
not\/} valid for the action including quarks.  Indeed, we can think of
it as multiplying different triality sectors containing $N$
quarks, (modulo $N$ for $SU(N)$) by different phases.  `Virtual' 
quark world-lines winding around the imaginary time circle are not left
invariant.  Indeed, with the implementation above, terms in the action 
that hop quarks from
$\tau =0 $ to $\tau = a$ are not invariant.
By redefining $\psi(\tau =a)$ by a phase you can move the changes in the
action to the next time slice, ... , but after winding around the
circle you will just arrive back where you started.

This formal argument based on confinement symmetry agrees, of course, 
with our previous
intuitive argument from consideration of conserved triality charge.  In the
pure glue theory there is a conserved flux that cannot be screened, an
associated symmetry, and a strict criterion of confinement; in the
theory with quarks all that structure is gone, and there is no strict
definition of confinement to distinguish it from screening.     

While there is no reason to expect the Polyakov loop strictly to
vanish at the onset of deconfinement (since deconfinement is, as I've
already belabored, an incoherent notion in this context), it remains a
perfectly respectable
observable.  One might expect that if there is a crossover from a
hadronic phase exhibiting pretty good confinement to a quark-gluon
phase with very poor confinement the Polyakov loop should take a
nose dive.  This is indeed the behavior that shows up in the numerical
simulation, as you see in Figure 10. 

\begin{figure}[!t]
\centerline{\psfig{figure=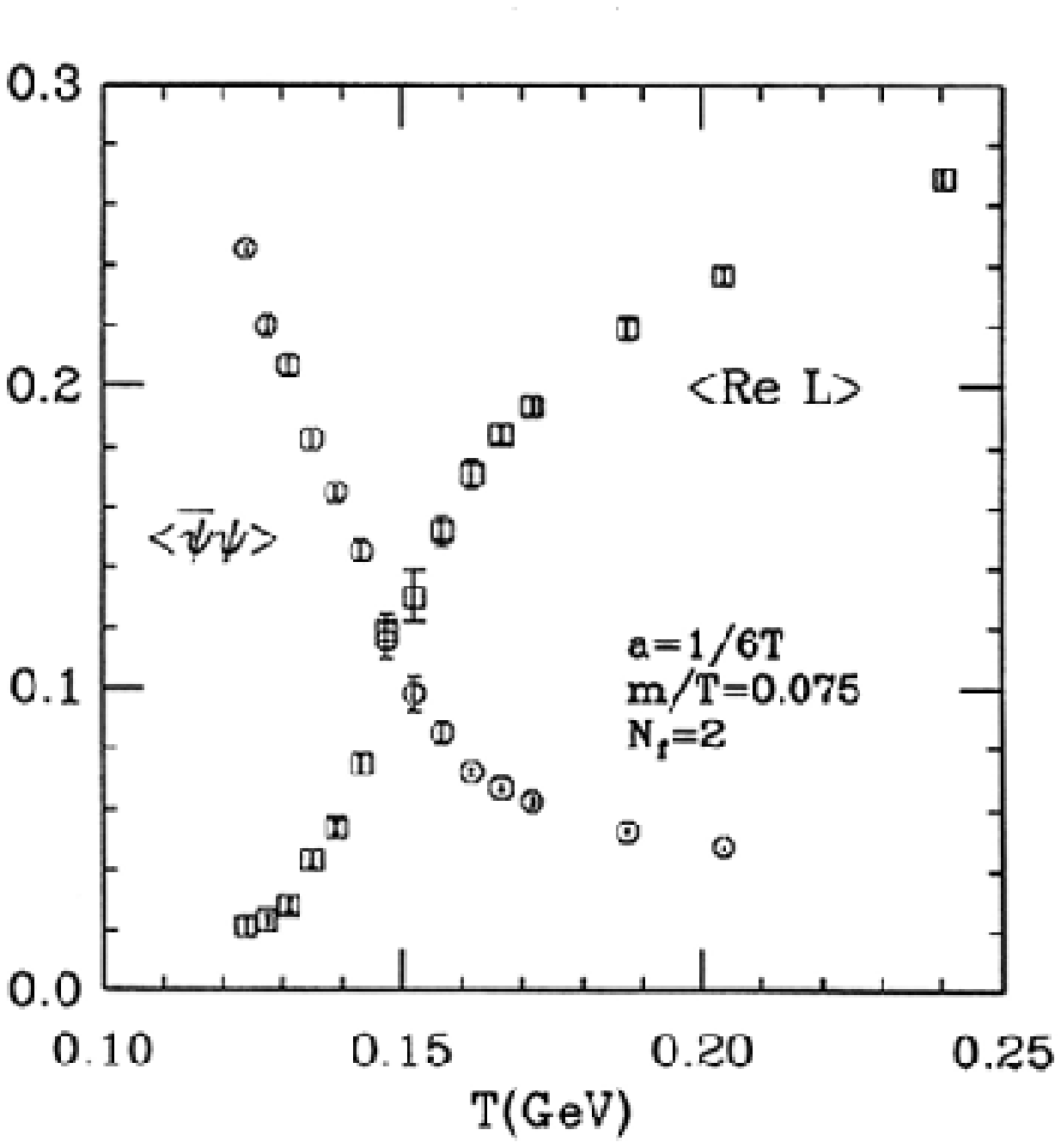,width=7cm}}
\vglue-.4in
\caption[]{Behavior of the
Polyakov loop $<L>$, and the chiral condensate $<\bar{\psi}\psi>$, in
2-Flavor QCD as functions of temperature. } 
\label{fig:working-bot.ps}
\end{figure}

The second is the chiral condensate.
While the notion of confinement gets fuzzy in the theory with quarks,
the notion of chiral symmetry breaking is perfectly
sharp (for massless quarks).  And it  
exhibits very interesting dynamical behavior as a function of
temperature.  The simplest measure of chiral symmetry breaking 
is simply the expectation value
$\langle \bar \psi \psi \rangle$.  This quantity is perfectly accessible to
lattice gauge theory.   In Figure 10, you see that this expectation
value does indeed take a dive.  In the simulation it does not reach
zero, because the quarks are not truly massless, but the possibility
of a smooth
decrease to zero at a finite value of $T$ is certainly suggested.
Studies with varying values of the quark masses, when extrapolated, further 
support this suggestion.

\section{Lecture 3: High-Temperature QCD: Phase Transitions}

\subsection{Yoga of Phase Transitions and Order Parameters}

Confinement, in the pure glue version of QCD (only), is a property we
can associate with a definite symmetry, that is valid at low
temperature but broken at high temperature.  Chiral symmetry, in the
versions of QCD with two or more massless quarks is, 
conversely, spontaneously
broken at low temperatures 
(at least if the number of quarks is not too large) but  
restored at high temperatures.  As emphasized by Landau, the presence
or absence of a symmetry is a sharp, objective question, which in any given
state of matter must have a yes or no answer.  And if the answer is yes in one
regime and no in another, passage from one regime to the other must be
accompanied by a sharp phase transition.  This situation 
is usually parameterized
by some appropriate order parameter, that transforms non-trivially
under the symmetry, and is zero on the unbroken side but non-zero on
the broken side. 

Phase transitions can occur without change of symmetry, or dynamical
reasons
-- we shall
see some simple examples below 
(where there is symmetry lurking just offstage).  
But by considering changes in
symmetry, which {\it must\/} be associated with phase transitions,  
and behaviors of order parameters we will be able to say quite a lot,
without doing any prohibitively difficult calculations.

\subsubsection{Second Order Transitions}

There are two broad classes of phase transitions, which have quite different
qualitative properties near the transition point.  First order
transition are characterized by a finite discontinuity in the generic
thermodynamic parameter -- {\it i.e}. basically in anything except the
free energy, which of course must be equal for the two phases at the
transition point.  Second-order transitions, on the other hand, are
characterized by continuous but nonanalytic behavior of thermodynamic
quantities.  

In Nature second order transitions are less common than first order
transitions, but they are especially interesting.  Near first
order transitions the two phases are simply `different', and are
described by distinct expressions for the free energy (in terms of
macroscopic variables).  There is a wholesale reorganization of
matter, even locally -- there are jumps in intensive variables.      
Near second order transitions that is not the case.  In a large but
finite volume, a first order transition point will be marked by rare but
sudden and
drastic jumps from one phase to the other, going over in the infinite
volume limit to hysteresis.  In the same circumstances, a second order
transition point will not exhibit any jumps, and the partition
function will be a perfectly analytic function.  

So how does the nonanalytic behavior arise?  It can only arise from
taking the infinite volume limit.  This, in turn, implies that for a
second order transition to occur there must be
low-energy fluctuations of arbitrarily long wave length, since it is
only such modes that can render the infinite volume limit subtle
(otherwise the free energy must ultimately become simply 
additive in the volume, for
large enough volumes).  In terms of static quantities, there 
must be a diverging correlation length.  In terms of particle physics,
there must be massless particles.  That is, 
if we quantized the modes under discussion, they would have massless quanta.

The hypothesis -- or quasi-theorem, as motivated above -- 
that nonanalytic behavior of thermodynamic quantities
near a second order transition must arise from the dynamics of massless
modes makes it possible, following Landau and Wilson, to make 
remarkably concrete and specific predictions about this behavior.  
The point is that it appears to be very difficult to construct consistent
theories of massless particles, unless one considers small
numbers of dimensions or large and exotic symmetry structures, that are 
inappropriate to the cases at hand.  So if we specify the desired
space dimension and symmetry we may find a unique theory of this kind, or
none at all.  Then the singular behavior of 
any possible second-order phase transitions with a given
dimensionality and symmetry will be uniquely determined, independent
of other details of the underlying microscopic theory.  This is the
hypothesis of {\it universality}. 

Universality makes it possible to make rigorous predictions for the 
behavior of 
complicated physical systems -- such as various versions
of QCD -- near second-order phase transitions by doing calculations in
much simpler models.

Now let me give a few words of orientation about the formal aspects of
such analyses, using the three dimensional Ising model as a prototype.
We are interested in the singular behavior of thermodynamic functions
near a possible second-order phase transition, where the magnetization
decreases from a non-zero value to zero.  The relevant low-energy,
long-wavelength modes are gradual changes in the local average of the
magnetization.  Since we are working at long wavelengths it is
appropriate to coarse grain, so the magnetization is described by a
real three-dimensional scalar field $\phi(x)$.  We are interested in
the singularity of the partition function induced by fluctuations of
$\phi(x)$.  To find it, we need to construct the appropriate
``universal'' theory based on $\phi(x)$.

We want this theory to be describing fluctuations that are small in
magnitude and long in wavelength, so we should use a Lagrangian with
the smallest possible powers of $\phi(x)$ and of derivatives.  
Of course we need a quadratic term with two derivatives to get any
non-trivial spatial behavior at all.  To give this term its chance to
shine, we will also need to put the mass equal to zero, since
a mass
term would always dominate the derivative term at long wavelength.
Since we have $\phi \rightarrow - \phi$ symmetry, the next possibility
is a $\phi^4$ term.  So our trial `Lagrangian' (to be interpreted as
$H/T$, the Hamiltonian divided by the temperature, in statistical
mechanical language) is  
\begin{equation}
\label{mela}
\int d^3 x {\cal L} ~ 
= ~ \int d^3 x ( \partial \phi )^2 + \lambda \phi^4 ~.
\end{equation}
Now if we count dimensions we see that for the Lagrangian to be
dimensionless $\phi$ must have mass dimension $1/2$, and so $\lambda$
must have mass dimension 1.   According to naive dimensional
analysis,  therefore, we are not getting a
scale-invariant theory.

We know, however, that interacting field theories contain another
source of non-trivial scaling behavior.  Because there are an infinite
number of degrees of freedom, we must regulate the theory, and define
a renormalized coupling at some finite momentum scale, which we fix to
a physical value independent of the cutoff.   Then in favorable cases
we will get
cutoff-independent answers, in terms of the renormalized coupling, 
as we take the cutoff to infinity.  We can get a scale invariant
theory from this set-up if the bare coupling $\lambda_b (\Lambda)$ 
one needs to insure a
fixed renormalized coupling scales as $\lambda_b \propto
\Lambda^{1\over 2}$.  The proportionality constant is then our
dimensionless parameter. 

That reasoning is quite abstract, so it is informative to consider the
situation also from a radically different perspective, due to Wilson
and Fisher.  We consider, formally, extending the theory to
$4-\epsilon$ dimensions.  It is of course problematic to construct a
field theory in non-integer dimensions, but we can perfectly well
continue the perturbation theory integrals, which ought to be adequate
if the relevant coupling turns out to be small.  Now in $4-\epsilon$
dimensions the same counting as above shows that the mass dimension of
$\lambda$ is $\epsilon$.  Thus as we rescale the momentum at which the
coupling is defined there are two sources of variation.  One is simple
classical dimensional analysis, which tends to make the $\phi^4$ more
relevant at small momenta; the other is the running due to
fluctuations (loops), familiar in its quantum version from QED or QCD,
where it is interpreted as vacuum polarization due to virtual
particles.  Of course here we are concerned with classical
fluctuations, but the equations are just the same.  Since $\phi^4$ in
4 dimensions is not asymptotically free, the two terms governing
renormalization toward the infrared go as
\begin{equation}
{d\tilde \lambda(p) \over dt} ~=~ 
\tilde \lambda \epsilon - b \tilde \lambda^2~,
\end{equation}
here $t = \ln (p_0 / p)$, with $p_0$ some reference momentum, 
and $\tilde \lambda \equiv \lambda p^{-\epsilon}$, and the equation is valid 
for $p << \Lambda$ the cutoff and 
small $\lambda$ and $\epsilon$.
$b$ is a calculable {\it positive\/} number.   So as 
$t \rightarrow \infty$
$\tilde \lambda \rightarrow \epsilon/b$, 
which is indeed small for small
enough $\epsilon$.  We say there is a fixed-point coupling with this
value.
With $\Lambda$ fixed, we approach 
a scale invariant theory for $p << \Lambda$.  The funny dependence of
the coupling $\lambda$ (not $\tilde \lambda$!)
extrapolated to momenta approaching the  
cutoff -- what we would call the bare coupling --
on the cutoff $\Lambda$ itself is just what we anticipated before, on
abstract grounds.      

This construction makes it plausible that there can be a
scale-invariant theory, but also makes it clear that this theory will
not be easy to find in three dimensions, where the fixed-point 
coupling cannot be
small.  One approach, which works remarkably well, is to calculate
around $4-\epsilon$ dimensions and extrapolate to $\epsilon =1$.  
Another is to work directly in three dimensions, calculate to high orders
in perturbation theory, and join on to the form at high orders, which
is known from sophisticated semiclassical techniques.  Finally, one
can simply simulate the theory directly numerically.  Any of these
techniques would be prohibitively difficult to use in high temperature
QCD directly, but thanks to universality we can get rigorous
quantitative answers
(to carefully selected questions, of course!) using much simpler models.

\subsubsection{First Order Transitions}

An important side-benefit of the analysis of how second-order
transitions arise is that it alerts us to cases where this cannot
occur.  In the Ising model analysis, we found the scale invariant
theory could arise when the mass parameter associated with the
magnetization vanishes.  Since we expect the effective mass parameter
to be a function of temperature, $m^2 = m^2(T)$, it is reasonable to
expect that this can happen 
at one particular value of the temperature.  So
Eqn.~\ref{mela} is a special case of the embedding set of Lagrangians
\begin{equation}
\int d^3x {\cal L} ~=~ \int d^3x (\partial \phi)^2 + m^2(T) \phi^2 +
\lambda \phi^4 ~
\end{equation}
describing the dynamics of the fluctuating magnetization not only
exactly at, but also near, the critical transition temperature.  This
is reasonable from another point of view as well: when $m^2(T)<0$
a non-zero expectation value for $\phi$ will be 
preferred, whereas for $m^2(T) > 0$ the expectation should vanish.

Now if we put the system in an external magnetic field, breaking the $\phi
\rightarrow - \phi$ symmetry, then $\phi$ and  $\phi^3$ terms are
allowed.  Let's shift away the $\phi$ term.  There will still be a
$T$-dependent $m^2$, and it can go through zero.  But now that will
generally not give rise to a second-order transition, because in the presence
of a small $\phi^3$ term the expectation value will jump to a new minimum at a
non-zero (positive) value of $m^2$.  The special case where the cubic
term vanishes simultaneously with $m^2$ 
can be accessed only if there is another control parameter available,
in addition to the temperature.  Then one has a so-called tricritical
point.  In the phase plane, the tricritical point appears as the
terminus of a line along which there are weaker and weaker first-order
transitions.

Even if the mean-field analysis allows a
second-order transition, there will not be one if there is no suitable
scale-invariant theory to represent the universality class.  The mean
field analysis ignored fluctuations, but as we have learned 
these are vital.  
In our discussion above, we saw that 
in order to construct the scale-invariant theory we needed to have a
simple, finite limiting behavior of the effective coupling under
renormalization group transformations toward the infrared.  
If there is no such limiting
behavior, there cannot be a second-order transition.  The physical
interpretation of this outcome is simply that in such cases
the fluctuations have grown out of
control, resulting in a catastrophic rearrangement of the state -- a
first-order transition.  Such an eventuality is, for obvious reasons, 
called a fluctuation driven first-order transition.     

Since they are marked by finite discontinuities, first-order
transitions are robust against small perturbations.  Thus if we have a
symmetry and an order parameter, whose change from a
non-zero to a zero value forces the existence of a first-order
transition according to either of the mechanisms I've just discussed,
there will still be a first-order transition even if the symmetry is
intrinsically 
slightly broken.  There will be no strict order parameter, and thus we
would not have been able to predict the necessity of a transition
without referring to the nearby, unbroken variant of the theory.

Each and every one of the theoretical phenomena I have mentioned in
these orienting sections plays a significant role in understanding the phase
structure of QCD!

\subsection{Application to Glue Theories}


Let's recall the basic facts.
When there are no quarks at all, then there is the possibility of
a true confinement-deconfinement transition.  As we have discussed,
such a transition is
characterized by the Polyakov order parameter
\begin{equation}
\label{aa}
\langle L  \rangle ~ =~ \langle~
{\rm Tr}~\exp\{ i\int_0^\beta A^a_\tau \lambda^a d\tau \}~ \rangle  .
\end{equation}
Here the expectation value is taken over the thermal ensemble of field
configurations periodic in imaginary time $\tau$ with period $\beta =
1/T$, and $\lambda$ is the representation matrix for the fundamental
representation.  This loop inserts quark quantum numbers into the
ensemble.  In the pure glue theory, the operator inserts a flux that
cannot be screened, and alters the state by an irreducible amount out
to infinity.  This costs a finite energy per unit volume, and
therefore infinite energy altogether.  The expectation value of the
loop would therefore be expected to vanish in the confined phase,
while it acquires a non-zero value in the unconfined phase.  $L$ is
multiplied by the appropriate complex root of unity when an element of
the center of the gauge group is applied to the state.  Thus symmetry
under this discrete group (triality for color $SU(3)$, diality for
$SU(2)$) is broken in the unconfined phase expected to exist at high
temperature.  

Since there is a simple order parameter with
well-defined symmetry properties, one can entertain the possibility of
a second order transition.  Indeed there does seem to be a second
order transition for $SU(2)$, in the universality class of the
(inverted) 3d Ising model.  `Inverted' refers to the features that
whereas in the Ising model the $Z_2$ symmetry is broken at low
temperature, but restored at high
temperature, the confinement $Z_2$ symmetry is valid at low
temperature but broken at high temperature.  This means the $m^2(T)$
goes through zero in the opposite direction, but of course one still 
has the
same universal theory at the critical point and a simple
correspondence away from it.

However for $SU(3)$ the appropriate model is different.  It is
something called 
the 3-state Potts model.  In the field-theoretic version of that
model we must use a complex scalar field $\phi$ invariant under
$\phi\rightarrow \omega \phi$, with $\omega$ the cube root of unity,
to implement the symmetry.  With such a field, cubic terms of the type
\begin{equation}
\Delta {\cal L} ~=~ \kappa (\phi^3 + \phi*^3 )
\end{equation}
are allowed.  The
existence of a cubic invariant implies that the
transition will be first order.

Both these predictions prove to be
true in large-scale numerical simulations of pure glue QCD.

\subsection{Application to Chiral Transitions}


Again, let's quickly recall the basics.
With dynamical quarks there is no longer a confinement symmetry,
but if we have $f$ flavors of massless quarks there is an additional
symmetry under chiral transformations in the group $SU(f)_L \times
SU(f)_R \times U(1)_B$ of independent special unitary rotations of the
left- and right-handed fields, together with the overall vector baryon
number symmetry.  (The additional apparent axial baryon number
symmetry, present at the classical level, is violated in quantum theory
by the anomaly, as discussed earlier.) 
This chiral
symmetry is believed on good grounds to break spontaneously down to
vector $SU(f)\times U(1)$ at low temperatures; and to be restored at
sufficiently high temperatures.  Of course for $f=1$ the chiral
symmetry is vacuous; but for $f\geq 2$ there is a phase transition
associated with restoration of chiral symmetry.  Since there is a
simple order parameter for this phase transition -- namely, for
example, the expectation value of the quark bilinear
\begin{equation}
\label{ab}
M^i_j ~=~ \langle  {\bar q_L}^i {q_R}_j \rangle
\end{equation}
-- one may again inquire concerning the possibility of a second order
transition.

\subsubsection{Formulation of Models}

In order to describe a possible second-order transition
quantitatively, we must try to find a tractable model in the
same universality class.  For the chiral order parameter
Eqn.~\ref{ab} the relevant symmetries are independent unitary transformations
of the left- and right-handed quark fields, under which

\begin{equation}
\label{ba}
M~\rightarrow~ U^\dagger M V ~.
\end{equation}
These transformations generate an
$SU(f)_L\times SU(f)_R \times U(1)_V $ symmetry, after the
anomaly in the axial baryon number current is taken into account.
At the phase transition, the true symmetry is broken to
$SU(f)_{L+R} \times U(1)_V$.

To describe the critical behavior, it is sufficient to retain the
degrees of freedom which develop long-range fluctuations at the
critical point.  It is natural to assume that these are associated
with long-wavelength variations in the order parameter, whose
magnitude is small and whose variations within the vacuum manifold
therefore 
cost little energy near the transition.  Thus the most plausible
starting point for analyzing the critical behavior of a possible
second-order phase transition in QCD is the
Landau-Ginzburg free energy
\begin{equation}
\label{bb}
{\cal F} = {\rm tr}~\partial_iM^\dagger \partial_i M
    ~+~ \mu^2{\rm tr}~M^\dagger M ~+~ \lambda_1{\rm tr}~(M^\dagger M)^2
    ~+~ \lambda_2({\rm tr}~ M^\dagger M )^2 ~.
\end{equation}
Here $\mu^2$ is the temperature-dependent renormalized (mass)$^2$,
which is negative below and positive above the critical point, while
$\lambda_1$ and $\lambda_2$ parameterize the strength of the quartic
couplings and are supposed to be smooth at the transition.  The
symmetry breaking pattern we want is $M \propto {\bf 1}$ below the
transition, which is what we shall find at the minimum of the
potential, within a range of positive $\lambda_1$ and $\lambda_2$.

Actually Eqn.~\ref{bb} is not quite what we want.  It has a full
$U(f)\times U(f)$ symmetry, which breaks down to $U(f)$.  Thus it
contains a massless Nambu-Goldstone boson for the axial baryon number
symmetry, which is not present in the microscopic theory (QCD)
we are trying to model.

For $f=2$ one can solve this problem very neatly by implementing the
symmetry in a slightly different way.  Instead of general complex
matrices, which in this case form a reducible representation of the
chiral symmetry, we can restrict ourselves to unitary matrices with
positive real determinant.  This restriction on $M$ is consistent with the
transformation law Eqn.~\ref{ba} as long as $U$ and $V$ have equal
phases, but not if the phases are unequal.  Thus the unwanted axial
$U(1)$ symmetry is indeed removed.  An important point is that for
$2\times 2$ matrices the condition of being a multiple of a unitary
matrix is a linear condition, so that we can enforce it while
remaining within the domain of renormalizable field theories.
(Nothing like this is true for larger matrices.)  It is very
convenient to parameterize the $2\times 2$ matrices in question in
terms of four real parameters $(\sigma,\vec \pi)$ and the Pauli
matrices as
\begin{equation}
\label{bc}
M~=~ \sigma + i\vec\pi \cdot \vec\tau~.
\end{equation}
In this way, we arrive back at the original model
of Gell-Mann and Levy.

After this pruning, the order parameter variables that have been
retained have the quantum numbers of the scalar isoscalar density
$\langle {\bar q}^iq_i \rangle $ and the pseudoscalar isovector
densities $\langle {\bar q}^x \gamma_5 \vec \tau q \rangle$.
Furthermore, the two {\it a priori\/} possible quartic couplings
$\lambda_1, \lambda_2$ are not independent --  a fact which proves
to be very significant. In fact the model boils down to the
theory of a four-component vector $\phi \equiv (\sigma, \vec \pi)$
in internal space, that is to say the standard $O(4)$ invariant
$n = 4$ ``Heisenberg magnet''.  For smaller numbers of components,
this sort of model is a much-studied model for the critical
behavior of magnets, with the vector of course representing the
magnetization (or staggered magnetization).

For larger values of $f$ the trick discussed above is no longer
available.  For $f = 3,4$ one can break the unwanted $U(1)$
symmetry by adding an additional determinantal interaction
${\rm det} M$.  Beyond $f=4$ that too is not entirely
satisfactory, because it takes
us outside the framework of renormalizable theories
in four dimensions (which are suitable starting points for
the construction of critical theories, using the
$\epsilon$ expansion).  However, as will soon appear, in the
present context it is almost certainly academic anyway.

\subsubsection{Fixed points}

Now we may search for second-order transitions within each
model, by the standard method of looking for infrared stable
fixed points of the renormalization group.  The vector
model has been studied in great depth, for arbitrary $n$.
The existence of a fixed point has been established by detailed
analysis of perturbation theory -- taken to high orders and
supplemented with estimates of the asymptotic behavior directly
in three dimensions -- directly in three dimensions, and also
{\it via\/} the $\epsilon$ expansion.  
The calculated critical exponents have been successfully compared with
appropriate experiments, for $n \leq 3$.  
Thus there can be no serious
doubt that there is a model for a second-order QCD chiral phase
transition for two massless quarks, consistent with the symmetry of
that theory.

On the other hand for $f\geq 3$ the structure of the renormalization
group for the appropriate model is more complicated -- there are
then two independent couplings, and so to speak more ways to go wrong.
In the lowest order of perturbation theory in the $\epsilon$ expansion,
the calculations can be done quite simply.
They indicate that
there is a fixed point of the renormalization group, but that it is
{\it not\/} infrared stable.  In plain English, this means (if we
can trust the $\epsilon$ expansion!) that near the transition one
would naively identify by taking $\mu^2$ through zero in Eqn.~\ref{bb} $M$ is
actually subject to catastrophic fluctuations, that change
the structure of the problem qualitatively.  Thus the renormalization
group fails to identify a consistent model for a second order transition,
and indicates instead that the transition must be a
first order transition.
In important work
Gausterer and Sanielevici have
simulated the $f=3$ matrix models directly, both with and without the
determinantal interaction, to search for second-order transitions.
None was found; the transition is always first order.  These authors
also studied the $f~=~2$ model, and did find the expected second-order
transition in that case.  Thus the expectations drawn from simple
$\epsilon$ expansion analysis appear to be vindicated.  Direct
simulation of the $n=4$ magnet model, and comparison with $f~=~2$ QCD
using the translation dictionary discussed below, is also an
attractive possibility. 

Existing numerical calculations for QCD, 
are consistent with the prediction 
that the proposed
models for the universality classes of possible QCD chiral phase
transitions are appropriate, and that there is a big difference between
the nature of the chiral transition in QCD for $f~=~2$ and for larger
$f$, with the former being second order and the latter first order.

\subsection{Close Up on Two Flavors}

If we accept that 
there is a second order transition for $f~=~2$, we can derive 
many precise, testable consequences.

As we have just discussed the most plausible starting point for
analyzing the critical behavior of a second-order phase transition in
QCD is the Landau-Ginzburg free energy
\begin{equation}
\label{BB}
F = \int d^3 x \Bigl\lbrace {1\over 2}~\partial^i \phi^\alpha
     \partial_i \phi_\alpha
    ~+~ {\mu^2 \over 2} ~\phi^\alpha \phi_\alpha ~+~
    {\lambda \over 4}(\phi^\alpha \phi_\alpha)^2
    ~\Bigr\rbrace ~,
\end{equation}
of an $n=4$ component scalar field.  Here $\mu^2$ is the
temperature-dependent renormalized (mass)$^2$, which is negative below
and positive above the critical point, while $\lambda$ is the strength
of the quartic couplings and is supposed to be smooth at the
transition.  We neglect terms with higher powers of $\phi$ since $|
\phi |$ is small near the transition. The symmetry breaking pattern we
want is ${\cal M} \propto {\bf 1}$ (equivalently, $\langle \sigma
\rangle \neq 0; \langle \vec \pi \rangle = 0$) below the transition
which is indeed what we find at the minimum of the potential for
positive $\lambda$.  This model has been studied in depth for
arbitrary $\eta$ and spatial dimension $d$, and the existence of an
infrared stable fixed point of the renormalization group for $\eta=4$,
$d=3$ is firmly established.
Hence, it is a model for a
second order QCD chiral phase transition for two massless quarks.
 
When the free energy Eqn.~\ref{BB} is written in terms of $\sigma$ and $\vec
\pi$ it looks much like the original model of Gell-Mann and Levy.
But there are two changes: there are no nucleon fields and
only three (spatial) dimensions.  These two changes reflect an
important distinction.
We are only proposing Eqn.~\ref{BB}
as appropriate near the second order phase transition point.  This is
because it is only there that we can appeal to universality -- the
long-wavelength behavior of the $\sigma$ and $\vec \pi$ fields is
determined by the infrared fixed point of the renormalization group,
and microscopic considerations are irrelevant to it.  In Euclidian
field theory at finite temperature, the integral over $\omega$ of zero
temperature field theory is replaced by a sum over Matsubara
frequencies $\omega_n$ given by $2n\pi T$ for bosons and $(2n+1)\pi T$
for fermions with $n$ an integer.  Hence, one is left with a Euclidian
theory in three spatial dimensions with massless fields from the $n=0$
terms in the boson sums and massive fields from the rest of the boson
sums and the fermion sums.  Hence, to discuss the massless modes of
interest at the critical point, Eqn.~\ref{BB} is sufficient.  We do not
need to introduce nucleon fields or constituent quark fields.

\subsubsection{Critical Exponents}
 
The most important universal properties of the second order transition
are the critical exponents, which we now define.

First, let us introduce
the reduced temperature $t~=~(T-T_c)/T_c$.
The exponents $\alpha$,  $\beta$, $\gamma$, $\eta$, and $\nu$
describe the singular
behavior of the theory with strictly zero quark masses as
$t \rightarrow 0$.
For the specific heat one finds
\begin{equation}
\label{da}
C(T) \sim |t|^{-\alpha} + {\rm less~ singular.}
\end{equation}
The behavior of the order parameter defines $\beta$.
\begin{equation}
\label{dab}
\langle | \phi |\rangle \sim  | t |^{\beta}~~ {\rm for}~ t<0~.
\end{equation}
$\eta$ and  $\nu$ describe the behavior of the correlation length $\xi$
where

\begin{equation}
\label{df}
 G_{\alpha \beta}(x)~ \equiv~
 \langle \phi(x)_\alpha \phi(0)_\beta \rangle
 -\langle \phi_\alpha \rangle \langle \phi_\beta \rangle
 ~\rightarrow~
 \delta_{\alpha \beta} {A \over |x|} \exp (-|x|/\xi) ~~{\rm 
at~large~distances.}
\end{equation}
$A$  is independent of $|x|$, but may depend on $t$.
The correlation length
exponent $\nu$ is defined by
\begin{equation}
\label{de}
\xi \sim |t|^{-\nu }~.
\end{equation}
Above $T_c$, where the correlation lengths are equal in the sigma and pion
channels,
the susceptibility exponent $\gamma$ is defined by
\begin{equation}
\label{deb}
\int d^3 x~G_{\alpha \beta}(x) \sim t^{-\gamma}.
\end{equation}
The exponent $\eta$ is defined through the
behavior of the Fourier transform of the correlation function:
\begin{equation}
\label{dea}
G_{\alpha \beta}(k\rightarrow 0) \sim k^{-2+\eta} ~.
\end{equation}
 
The last exponent, $\delta$ is related to the behavior
of the system in a small magnetic field $H$ which explicitly
breaks the $O(4)$ symmetry.  Let us first show that in a QCD context,
$H$ is proportional to a common quark mass $m_u = m_d \equiv m_q$.
This common mass term may be represented by a $2\times 2$ matrix ${\cal D}$
given by $m_q$ times the identity matrix.
We are now allowed to construct the free energy from invariants involving
both ${\cal D}$ and ${\cal M}$.  
The lowest dimension term linear in ${\cal D}$
is just ${\rm tr} {\cal M}^\dagger {\cal D} = m_q \sigma$,
which in magnet language
is simply the coupling of the magnetization to an external field $H \propto
m_q$.  In the presence of an external field, the order parameter is
not zero at $T_c$.  In fact,
\begin{equation}
\label{dec}
\langle | \phi |\rangle (t=0,H\rightarrow 0) \sim H^{1/\delta}~.
\end{equation}
 
The six critical exponents defined above are related by four scaling
relations. 
These are
\begin{eqnarray}
\label{db}
\alpha~&=~2-d\nu  \nonumber  \\
\beta~&=~{\nu \over 2} ( d - 2 + \eta ) \nonumber \\
\gamma~&=~(2-\eta )\nu  \nonumber \\
\delta~&=~{d + 2 - \eta \over d - 2 +\eta}~.  
\end{eqnarray}
We therefore need values for $\eta$ and $\nu$ for the four component
magnet in $d=3$. These were obtained in the remarkable work of
Baker, Meiron and Nickel,
who carried the perturbation theory to
seven-loop order, and used information about the behavior of
asymptotically large orders, and conformal mapping and Pad\'e
approximant techniques to obtain
\begin{eqnarray}
\label{dc}
\eta ~&=~~ .03 \pm .01 \nonumber \\
\nu  ~&=~~ .73 \pm .02 ~.
\end{eqnarray}
Using Eqn.~\ref{db}, the remaining exponents are
$\alpha = -0.19 \pm .06$, $\beta = 0.38 \pm .01$,
$\gamma = 1.44 \pm .04$ and $\delta = 4.82 \pm .05$.  Since $\alpha$
is negative there is a cusp in the specific heat at $T_c$, rather than a
divergence.

A more powerful result which includes Eqn.~\ref{dab} and Eqn.~\ref{dec} 
as special cases 
is the critical equation of state:
\begin{equation}
\label{du}
H~=~ M|M|^{\delta -1}\kappa_1 g(\kappa_2 t|M|^{-{1\over \beta}})
\end{equation}
in which here $H~\equiv~m$, 
$M~\equiv~\langle {\bar q} q \rangle = \langle |\phi| \rangle$, 
$t~\equiv~(T-T_c)/T_c$, 
$g$ is a universal function, and $\kappa_1$ and
$\kappa_2$ are non-universal constants.  This equation of state could,
of course, be compared directly with sufficiently 
accurate numerical simulations.

\subsubsection{Tricritical Point}

To this point I have been discussing a world with two massless
quarks, and so we have implicitly been taking the strange quark mass
to be infinite.  On the other hand, I argued earlier that if the
strange quark is massless, then the chiral phase transition is first
order.  Hence, as the strange quark mass is reduced from infinity to
zero, at some point the phase transition must change from second order
to first order.  This point is called a tricritical point.  There is
numerical evidence  that when the strange quark has its
physical mass, and the two light quarks are taken strictly massless,
the transition is second order, with the consequences discussed above.
However this conclusion was controversial for many years, and even now
may not be completely secure.  In any case the physical strange
quark mass is not drastically different from the critical value, so
let's indulge our idle curiosity a little further (this will pay off
shortly).

In a lattice simulation, the strange
quark mass can be tuned to just the right value to reach the
tricritical point.  Let's discuss the critical exponents
that would be observed in such a simulation.  
 
Let us consider the effect of adding a massive but not infinitely
massive strange quark to the two flavor theory.  This will not
introduce any new fields which become massless at $T_c$, and so the
arguments leading to the free energy Eqn.~\ref{BB} are still valid.  The only
effect of the strange quark, then, is to renormalize the couplings.
Renormalizing $\mu^2$ simply shifts $T_c$, as does renormalizing
$\lambda$ unless $\lambda$ becomes negative.  In that case, one can no
longer truncate the Landau-Ginzburg free energy at fourth order.
After adding a sixth order term, and keeping track of small light
quark masses, the free energy becomes
\begin{equation}
\label{sixth}
F = \int d^3 x \Bigl\lbrace {1\over 2}(\nabla \phi)^2 + {\mu^2 \over  2}\phi^2
+ {\lambda \over 4}(\phi^2)^2 + {\kappa \over 6}(\phi^2)^3 - H\sigma
\Bigr\rbrace ~.
\end{equation}
While for positive $\lambda$, $\phi^2$ increases continuously from zero
as $\mu^2$ goes through zero, for negative $\lambda$, $\phi^2$
jumps discontinuously from zero to $|\lambda |/(2\kappa)$ when
$\mu^2$ goes through $\lambda^2 /(4\kappa)$.  Hence, the phase transition
has become first order.  Thus at the value of $m_s$ where $\lambda = 0$,
the phase transition changes continuously from second order to first order.
 
The singularities of thermodynamic functions near tricritical points, like
the singularities near ordinary critical points, are universal.  Hence,
it is natural to propose  that QCD with two
massless flavors of quarks and with $T$ near $T_c$ and $m_s$ near its
tricritical value is in the universality class of the $\phi^6$ Landau-Ginzburg
model Eqn.~\ref{sixth} .  This model has been studied extensively.
Because the $\phi^6$ interaction is strictly renormalizable in three 
dimensions,
this model is much simpler to analyze than the $\phi^4$ model of the
ordinary critical point.  No $\epsilon$ expansion is necessary, and
the critical exponents all take their mean field values, up to
calculable logarithmic corrections.  Here I'll be content to show you
the mean field tricritical exponents.
 
In mean field theory, the correlation function in momentum space is simply
$G_{\alpha \beta}(k) = \delta_{\alpha \beta} (k^2 + \mu^2)^{-1}$.
Since $\mu^2 \sim t$, this gives the exponents $\eta =0$, $\gamma =1$
and $\nu = 1/2$.  To calculate $\alpha$ and $\beta$, we minimize
$F$ for $H=\lambda=\nabla \phi =0$, and find $\alpha = 1/2$ and
$\beta = 1/4$.
To calculate $\delta$, we minimize $F$ for $t=\lambda =\nabla \phi =0$
and find $\delta = 5$.
 
The result for the specific heat exponent $\alpha$
is particularly interesting, since it means that
the specific heat diverges at the tricritical point, unlike at the
ordinary critical point.
This means that whereas for $m_s$ large enough that the transition
is second order the specific heat $C(T)$ has a cusp but is finite
at $T=T_c$, as $m_s$ is lowered to the tricritical value $C(T_c)$
should increase since at the tricritical point it diverges.  This behavior
should be seen in future lattice simulations.
 
Finally, at a tricritical point there is one more
relevant operator than at a critical point, since two physical quantities
($t$ and $m_s$) must be tuned to reach a tricritical point.  Hence,
a new exponent $\phi_t$, the crossover exponent, is required.
For $\lambda \neq 0$, tricritical behavior
will be seen only for $|t|>t^*$, while for $|t|<t^*$, either ordinary
critical behavior or first order behavior (depending on the sign
of $\lambda$) results.  $t^*$ depends on $\lambda$ according to
\begin{equation}
\label{cross}
t^* \sim \lambda^{1/\phi_t}
\end{equation}
The mean field value of $\phi_t$ is obtained by minimizing the free
energy $F$ for $H=\nabla \phi =0$, and is $\phi_t = 1/2$.
These mean field tricritical exponents, $\alpha = 1/2$, $\beta = 1/4$,
$\gamma = 1$, $\delta = 5$, $\eta = 0$, $\nu = 1/2$, and $\phi_t = 1/2$
would describe the real
world if $m_s$ were smaller than it is, and will describe
future lattice simulations with $m_s$ chosen appropriately.

\subsubsection{Robustness}

If we now turn on small but non-zero $u$ and $d$ quark masses, the
first-order line will persist, but the second-order transition will
disappear, replaced by a crossover.  And there will still be a true
(tri)critical point, where the first-order line terminates!  It will
be in the universality class of a liquid-gas or Ising tricritical
point, rather than the asymptotically free $\phi^6$-type theory.

\subsection{A Genuine Critical Point! (?)}

In reality, of course, we cannot  vary the strange quark mass.  
If the strange quark mass is, as seems to be indicated, so large that
for massless $u$ and $d$ quarks we would have a second-order
transition, then after taking into account the non-zero mass of these
quarks we are left with only a crossover.  Moreover since the mass of
the physical pions is not so different from the critical temperature,
we cannot expect that correlation length ever gets very long.  Thus we
are left with a crossover, and perhaps not a very dramatic one.  This
is a definite prediction, though perhaps a disappointing one for
experimentalists, of the form
``nothing striking occurs'' (in the way of sharp phase transitions).  
But fortunately, as emphasized in recent work of Stephanov,
Rajagaopal, and Shuryak, that is not the end of the story.

Their basic insight is that although one cannot vary the strange quark
mass experimentally, there is another control parameter that one can
vary -- the chemical potential.   If the singularity in $T$ at
slightly unphysical values of $m_s$ and zero $\mu$ continues into a
singularity at the physical value of $m_s$ for slightly costive $\mu$,
we will be back in business, with a true physical (tri)critical point,
featuring diverging correlation lengths at a particular point in the $\mu,  
T$ plane.

\subsubsection{Subnuclear Boiling}

The same structure is also suggested by an entirely different line of
thought.  Consider now the behavior of two-flavor QCD 
as one varies the chemical
potential $\mu$ at {\it zero\/} temperature.  Models of the type we
will be considering in the next two lectures suggest that in the case
of two mass quarks there is a
first-order transition from a state of broken chiral symmetry at 
$\mu = 0$ to a state with chiral symmetry restored at large $\mu$.
Since the transition is first order, it remains as a first-order
transition in the presence of small symmetry-breaking perturbations
(the quark masses), despite the absence of an order parameter.  
This idea is very much in the spirit of the phenomenologically
successful MIT bag model, in which nucleons are pictured as droplets
of chiral symmetry restored phase embedded in a broken symmetry
vacuum.  Indeed, if true, it comes very close to providing a
microscopic justification of that model.  

Physically, the suggested transition represents a sort of subnuclear
boiling, to a phase in which quarks (or at least, as we shall see,
a subset of them) are liberated to behave as nearly massless, weakly
interacting particles.  

If in fact there is such a transition, it becomes interesting to
follow its fate as a function of temperature.  At small temperatures,
we shall still have a first-order transition at some value $\mu_c(T)$. 
We might expect that $\mu_c(T)$ decreases with $T$, since higher
temperature should favor the higher-entropy `boiled' state.  This
expectation can be convincingly justified on thermodynamic grounds.   
But we know that $\mu_c(T)$ cannot decrease to zero, 
since for $\mu =0 $ we have 
only a crossover.  Again, the
simplest possibility to reconcile the behaviors along the axes is to
suppose that the line of first-order transitions terminates at a
definite $(\mu_t, T_t)$.  The strength of the first-order transition
weakens as a function of temperature, until at $T_t$ the strict distinction
between hadron and quark phases has disappeared.  This $(\mu_t, T_t)$
is the ripe fruit of our theoretical labors: the prediction of 
a true (tri)critical point in
honest-to-god real world QCD, arrived at from
qualitative, but deeply rooted, theoretical considerations!

\subsubsection{Physical Signatures}

Stephanov, Rajagopal and Shuryak (SRS) have made quite specific and detailed
proposals for experimental signatures of the phase transition.  A
proper discussion of this would take us far afield, but a few comments
are in order.

In a heavy ion collision one produces a baryon number poor fireball in
the central region, but toward the fragmentation regions the baryon
number density increases.  As the different parts of the fireball
expand and cool, each part traces out a different history in the
$(\mu, T)$ plane.  There will be critical fluctuations and long
correlation lengths in a given region of phase space if -- and only if
-- material in the region has passed close close to $P = (\mu_t, T_t)$
during its history.  For that class of events and regions of rapidity
space, one can expect enhanced multiplicities of, and correlations
among, low-momentum pions.  These enhancements will have the
distinctive feature of being non-monotonic in the experimental control
parameters, since one can miss the critical point on either of two
sides.


The simplest observables to analyze are the event-by-event
fluctuations of the mean transverse momentum of the charged particles
in an event, $p_T$, and of the total charged multiplicity in an event,
$N$.  SRS calculate the magnitude of the effects of critical
fluctuations on these and other observables, making predictions which,
they hope, will allow experiments to find.  As a necessary prelude,
they analyze the contribution of noncritical thermodynamic
fluctuations.  They find  that NA49 data
(hep-ex/9904014) is consistent with the hypothesis that most of the
event-by-event fluctuation observed in the data is thermodynamic in
origin.  This bodes well for the detectability of systematic changes
in thermodynamic fluctuations near $P$.

As one example, consider the ratio of the width of the true
event-by-event distributions of $p_T$ to the width of the distribution
in a sample of mixed events.  SRS call this ratio $\sqrt{F}$.  NA49
has measured $\sqrt{F} = 1.002 \pm 0.0002$, which is consistent with
expectations for noncritical thermodynamic fluctuations.  A detailed
calculation suggests 
that critical fluctuations can increase $\sqrt{F}$ by 10 - 20\%,
fifty times the statistical error in the present measurement.  There
are other observables which are even more sensitive to critical
effects.  For example, a $\sqrt{F_{\rm soft}}$ defined by using only the
10\% softest pions in each event, might well be affected at the factor
of two level.  

NA49 data demonstrates very clearly that SPS collisions at $\sqrt{s}$ =
17 GeV {\it do not} freeze out near the critical point.  $P$ has not
yet been discovered.  The nonmonotonic appearance and then
disappearance (as $\sqrt{s}$ is varied) of any one of the signatures
of the critical fluctuations mentioned above, or others, would be strong
evidence for critical fluctuations.  If nonmonotonic variation is seen
in several of these observables, with the maxima in all signatures
occurring at the same value of $\sqrt{s}$, it would turn strong
evidence into an unambiguous discovery of the critical point.  The
quality of the present NA49 data, and the confidence with which we can
use it to learn that collisions at $\sqrt{s}$ = 17 GeV
17 GeV do not freeze out near the critical point make it plausible
that critical behavior, if present, could be discerned
experimentally. 
If and when 
the critical point $P$ is
discovered, it will appear prominently on the map of the phase diagram
featured in any future textbook of QCD.

\subsubsection{Wanted: Numerical Data}

The web of general arguments I have outlined above seems to me to make
it quite plausible that there is a true critical point in QCD.  We can
invoke the canonical yoga of second-order phase transitions to make
precise, quantitive predictions for the singular behavior of
thermodynamics near this point.  Arguments of this sort, however,
cannot address the non-universal but vital question of precisely where
$(\mu_t, T_t)$ is.  It is a great challenge to locate this point
theoretically, and of course such knowledge, even if approximate,
would greatly simplify the experimentalist search.

I believe it ought to be possible to locate the tricritical point
numerically.  For while it is notoriously difficult to deal with large
chemical potentials at small temperature numerically, there are good
reasons to be optimistic about high temperatures and relatively small
chemical potentials, which is our concern here.
Some combination of extrapolating from data taken at imaginary values
of the chemical potential, and/or using the real part of the fermion
determinant as the measure, may be adequate to sense the singularity,
especially if we dial the strange quark mass to bring it close to $\mu
= 0$ (and then extrapolate back to the physical strange quark mass).

If all these strands can be brought together, it will be a wonderful
interweaving of theory, experiment, and numerics.


\section{Lecture 4: High-Density QCD: Methods}

The behavior of QCD at high density is intrinsically interesting, as
the answer to the question: What happens to matter, if you keep
squeezing it harder and harder?  It is also directly relevant to the
description of neutron star interiors, neutron star collisions, and
events near the core of collapsing stars.  Also, one might hope to
obtain some insight into physics at ``low'' density -- that is,
ordinary nuclear density or just above -- by approaching it from the
high-density side.

\subsection{Hopes, Doubts, and Fruition}

Why might we anticipate QCD simplifies in the limit of high density?
A crude answer is: ``Asymptotic freedom meets the Fermi
surface.''   One might argue, formally, that the only external
mass scale characterizing the problem is the large chemical potential
$\mu$, so that if the effective coupling $\alpha_s(\mu)$ is small, as
it will be for $\mu \gg \Lambda_{QCD}$, where $\Lambda_{QCD} \approx
200$ Mev is the primary QCD scale, then we have a weak coupling
problem.  More physically, one might argue that at large $\mu$ the
relevant, low-energy degrees of freedom involve modes near the Fermi
surface, which have large energy and momentum.  An interaction between
particles in these modes will either barely deflect them, or will
involve a large momentum transfer.  In the first case we don't
care, while the second is governed by a small effective coupling.

These arguments are too quick, however.  The formal argument is
specious, if the perturbative expansion contains infrared divergences.
And there are good reasons -- two separate ones, in fact -- to
anticipate such divergences.

First, Fermi balls are generically unstable against the effect of
attractive interactions, however weak, between pairs near the Fermi
surface that carry equal and opposite momentum.  This is the Cooper
instability, which drives ordinary superconductivity in metals and the
superfluidity of He3.  It is possible to have an instability at
arbitrarily weak coupling because occupied pair states can
have very low energy, and they can all scatter into one another.  Thus
one is doing highly degenerate perturbation theory, and in such a
situation even a very weak coupling can produce significant
``nonperturbative'' effects.

Second, nothing in our heuristic argument touches the gluons.  To be
sure the gluons will be subject to electric screening, but at zero
frequency there is no magnetic screening, and infrared divergences do
in fact arise, through exchange of soft magnetic gluons.

Fortunately, by persisting along this line of thought we find a path
through the apparent difficulties.  Several decades ago Bardeen,
Cooper, and Schrieffer taught us, in the context of metallic
superconductors, how the Cooper instability is resolved. We can easily
adapt their methods to QCD.  In electronic systems only rather 
subtle mechanisms can generate an attractive effective interaction
near the Fermi surface, since the primary electron-electron
interaction is Coulomb repulsion.  In QCD, remarkably, it is much more
straightforward.  Even at the crudest level we find attraction.
Indeed, two quarks, each a color triplet, can combine to form a single
color antitriplet, thus reducing their total field energy.

The true ground state of the quarks is quite different from the naive
Fermi balls.  It is characterized by the formation of a coherent
condensate, and the development of an energy gap.  The condensation,
which is energetically favorable, is inconsistent with a magnetic color
field, and so such weak magnetic color 
fields are expelled.  This is the color version of the famous
Meissner effect in superconductivity, which is essentially identical
to what is known as the Higgs phenomenon in particle physics.
Magnetic screening of gluons, together with energy gaps for quark
excitations, remove the potential sources of infrared divergences
mentioned above.  Thus we have good reasons to hope that a weak
coupling -- though, of course, nonperturbative -- treatment of the
high density state will be fully consistent and accurate.

The central result in recent developments is that this program can
be carried to completion rigorously in QCD with sufficiently many
(three or more) quark species.  Thus the more refined, and
fully adequate, answer to our earlier question is: ``Asymptotic
freedom meets the BCS groundstate.''  Together, these concepts can
render the behavior of QCD at asymptotically high density calculable.

\subsection{Another Renormalization Group}

To sculpt the problem, begin by assuming weak coupling, and focus on
the quarks.  Then the starting point is Fermi balls for all the
quarks, and the low-energy excitations include states where some modes
below the nominal Fermi surface are vacant and some modes above are
occupied.  The renormalization group, in a generalized sense, is a
philosophy for dealing with problems involving nearly degenerate
perturbation theory.  In this approach, one attempts to map the
original problem onto a problem with fewer degrees of freedom, by
integrating out the effect of the higher-energy (or, in a relativistic
theory, more virtual) modes.  Then one finds a new formulation of the
problem, in a smaller space, with new couplings.  In favorable cases
the reformulated problem is simpler than the original, and one can go
ahead and solve it.

This account of the renormalization group might seem odd, at first
sight, to high-energy physicists accustomed to using asymptotic
freedom in QCD.  That is because in traditional perturbative QCD one
runs the procedure backward.  When one integrates out highly virtual
modes, one finds the theory becomes more strongly coupled.  Simplicity
arises when one asks questions that are somehow inclusive, so that to
answer them one need not integrate out very much.  It is then that the
microscopic theory, which is ideally symmetric and constrained,
applies directly.  So one might say that the usual application of the
renormalization group in QCD is fundamentally negative: it informs us
how the fundamentally simple theory comes to look complicated at low
energy, and helps us to identify situations where we can avoid the
complexity.
\begin{figure}
\begin{center}
\hspace{0.3cm}
\vspace{0.5cm}
\epsfxsize=8cm
\epsffile{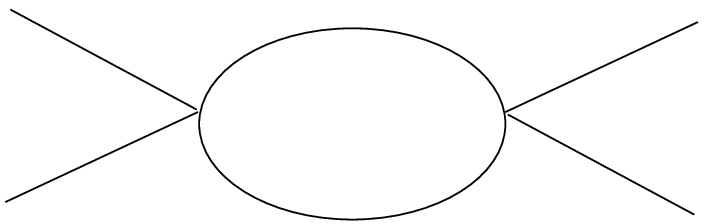}
\end{center}
\vspace*{-1cm}
\caption{Graph contributing to the renormalization of four-fermion
couplings. }
\end{figure}

Here, although we are still dealing with QCD, we are invoking quite a
different renormalization group, one which conforms more closely to
the Wilsonian paradigm.  We consider
the effect of integrating out modes whose energy is within the band
$(\epsilon, \delta \epsilon)$ of the Fermi surface, on the modes of
lower energy.  This will renormalize the couplings of the remaining
modes, due to graphs like those displayed in Figure 11.  In addition
the effect of higher-point interactions is suppressed, because the
phase space for them shrinks, and it turns out that only four-fermion
couplings survive unscathed (they are the marginal, as opposed to
irrelevant, interactions).  Indeed the most significant interactions
are those involving particles or holes with equal and opposite three
momenta, since they can scatter through many intermediate states.  For
couplings $g_\eta$ of this kind we find
\begin{equation}
\label{fermiRG}
{d g_\eta \over d \ln \delta } = \kappa_\eta g_\eta^2. 
\end{equation}
Here $\eta$ labels the color, flavor, angular momentum, ...  channel
and in general we have a matrix equation --
but let's keep it simple, so $\kappa_\eta$ is a positive number.
Then Eqn.~\ref{fermiRG} is quite simple to integrate, and we have
\begin{equation}
\label{ked}
{1\over g_\eta (1) } - {1\over g_\eta(\delta ) } =  \kappa_\eta \ln \delta.
\end{equation}
Thus for $g_\eta(1)$ negative, corresponding to attraction, $|g_\eta
(\delta)|$ will grow as $\delta \rightarrow 0$, and become singular
when
\begin{equation}
\label{let}
\delta =  e^{1\over \kappa_\eta g_\eta (1)}.
\end{equation}
Note that although the singularity occurs for arbitrarily weak
attractive coupling, it is nonperturbative.

\subsection{Pairing Theory}

The renormalization group toward the Fermi surface helps us identify
potential instabilities, but it does not indicate how they are
resolved.  The great achievement of BCS was to identify the form of
the stable ground state the Cooper instability leads to.  Their
original calculation was variational, and that is still the most
profound and informative approach, but simpler, operationally
equivalent algorithms are now more commonly used.  I will be very
sketchy here, since this is textbook material.

The simplest and most beautiful results, luckily, occur in the version
of QCD containing three quarks having equal masses.  I say luckily,
because this idealization applies to the real world, at densities so
high that we can neglect the strange quark mass (yet not so high that
we have to worry about charmed quarks).  Until further notice, I'll be
focusing on this case.

Most calculations to date have been based on model interaction
Hamiltonians, that are motivated, but not strictly derived, from
microscopic QCD.  They are chosen as a compromise between realism and
tractability.  For concreteness I shall here follow, and
consider
\begin{equation}
\label{modelH}
\begin{array}{ll}
H &= \dsp\int d^3x\, \psib(x) (
\nabla\slash - \mu\ga_0) \psi(x) + H_I, \\
H _I &= K \dsp\sum_{\mu, A}\int d^3x\, {{\cal F}}
 \psib(x) \ga_\mu T^A \psi(x)\ \psib(x) \ga^\mu T^A \psi(x)\ 
\end{array}
\end{equation}
Here the $T^A$ are the color $SU(3)$ generators, so the quantum
numbers are those of one-gluon exchange.  However instead of an honest
gluon propagator we use an instantaneous contact interaction, modified
by a form-factor $\cal F$.  $\cal F$ is taken to be a product of
several momentum dependent factors $F(p)$, one for each leg, and to
die off at large momentum.  One convenient possibility is $F(p) =
(\lambda^2/(p^2 + \lambda^2))^\nu$, where $\lambda$ and $\nu$ can be
varied to study sensitivity to the location and shape of the cutoff.
The qualitative effect of the form-factor is to damp the spurious
ultraviolet singularities introduced by $H_I$; microscopic QCD, of
course, does have good ultraviolet behavior.  One will tend to trust
conclusions that do not depend sensitively on $\lambda$ or $\nu$. In
practice, one finds that the crucial results -- the form and magnitude
of gaps -- are rather forgiving.

Given the Hamiltonian, we can study the possibilities for symmetry
breaking condensations.  The most favorable condensation possibility
so far identified is of the form
\begin{equation}
\label{Cond}
\begin{array}{lll}
~\langle q^{i\alpha}_{La} (p) q^{j\beta}_{Lb} (-p) \rangle ~=~ 
-\langle q^{i\alpha}_{Ra} (p) q^{j\beta}_{Rb} (-p) \rangle ~=~ 
\epsilon^{ij} (\kappa_1(p^2) \delta^\alpha_a \delta^\beta_b +
\kappa_2(p^2) \delta^\alpha_b \delta^\beta_a)~. 
\end{array}
\end{equation}
Here we encounter the
phenomenon of color-flavor locking.  The ground state contains
correlations whereby both color and flavor symmetry are spontaneously
broken, but the diagonal subgroup, which applies both transformations
simultaneously, remains valid.
There are several good reasons to think that condensation of this form
characterizes the true ground state, with lowest energy, at asymptotic
densities.  It corresponds to the most singular channel, in the
renormalization group analysis discussed above.  It produces a gap in
all channels, and is perturbatively stable, so that it is certainly a
convincing local minimum.  It resembles the known order parameter for
the B phase of superfluid He3.  And it beats various more-or-less plausible
competitors that have been investigated, by a wide margin.

Given the form of the condensate, one can fix the leading functional
dependencies of $\kappa_1(p^2, \mu)$ and $\kappa_2(p^2, \mu)$ at weak
coupling by a variational calculation.  For present purposes, it is
adequate to replace all possible contractions of the quark fields in
Eqn.~\ref{modelH} having the quantum numbers of Eqn.~\ref{Cond} with their
supposed expectation values, and diagonalize the quadratic part of the
resulting Hamiltonian.  The ground state is obtained, of course, by
filling the lowest energy modes, up to the desired density.  One then
demands internal consistency, {\it i.e}. that the postulated
expectation values are equal to the derived ones.  Some tricky but
basically straightforward algebra leads us to the result
\begin{equation}
\label{pDependence}
\Delta_{1,8} (p^2) = F(p)^2 \Delta_{1,8} 
\end{equation}
where $\Delta_1$ and $\Delta_8$ satisfy the coupled gap equations
\begin{equation}
\label{coupG}
\begin{array}{ll}
 \Delta_8 + {1\over 4} \Delta_1 &  = {16\over 3}K G(\Delta_1) \\ 
\noalign{\medskip}
{1\over 8} \Delta_1  &  =  {16\over 3}K G(\Delta_8) 
\end{array}
\end{equation}
where we have defined
\begin{equation}
\label{coupGaps}
G(\Delta) = -{1\over 2}\sum_{\bk}\Biggl\{
  {F(k)^4 \Delta\over \sqrt{ (k-\mu)^2 + F(k)^4 \Delta^2}}  
+ {F(k)^4 \Delta\over \sqrt{ (k+\mu)^2 + F(k)^4 \Delta^2} }\Biggr\}
\end{equation}
and
\begin{equation}
\label{dok}
\begin{array}{ll}
\kappa_1(p^2) & = {1\over 8K}(\Delta_8(p^2) + {1\over 8} \Delta_1(p^2)) \\
\noalign{\medskip}
\kappa_2(p^2) & = {3\over 64K} \Delta_1(p^2).
\end{array}
\end{equation}
The $\Delta$ are defined so that $F(p)^2 \Delta_{1,8}(p^2)$ are the
gaps for singlet or octet excitations at 3-momentum $p$.   
Eqn.~\ref{coupG}  must be solved numerically.  

Finally, to obtain quantitative estimates of the gaps, we must
normalize the parameters of our model Hamiltonian.  One can do this
very crudely by using the model Hamiltonian in the manner originally
pioneered by Nambu and Jona-Lasinio, that is as the basis
for a variational calculation of chiral symmetry breaking at zero
density.  The magnitude of this chiral condensate can then be fixed to
experimental or numerical results.  In this application we have no
firm connection between the model and microscopic QCD, because there
is no large momentum scale (or weak coupling parameter) in sight.
Nevertheless a very large literature following this approach
encourages us to hope that its results are not wildly wrong,
quantitatively.  Upon adopting this normalization procedure, one finds
that gaps of order several tens of Mev near the Fermi surface are
possible at moderate densities.

While this model treatment captures major features of the physics of
color-flavor locking, with a little more work it is possible to do a
much more rigorous calculation, and in particular to normalize
directly to the known running of the coupling at large momentum.  I'll 
now sketch .

\subsection{Taming the Magnetic Singularity}

A proper discussion of the fully microscopic calculation is necessarily quite
technical, and would be out of place here, but the spirit of the thing 
-- and one of the most striking results -- can be conveyed simply. 

When retardation or relativistic effects are important a Hamiltonian
treatment is no longer appropriate.  One must pass to Lagrangian and
graphical methods.  (Theoretical challenge: is it possible to
systematize these in a variational approach?)  The gap equation
appears as a self-consistency equation for the assumed condensation,
shown graphically in Figure 12.
\begin{figure}
\begin{center}
\hspace{0.3cm}
\vspace{0.5cm}
\epsfxsize=10cm
\epsffile{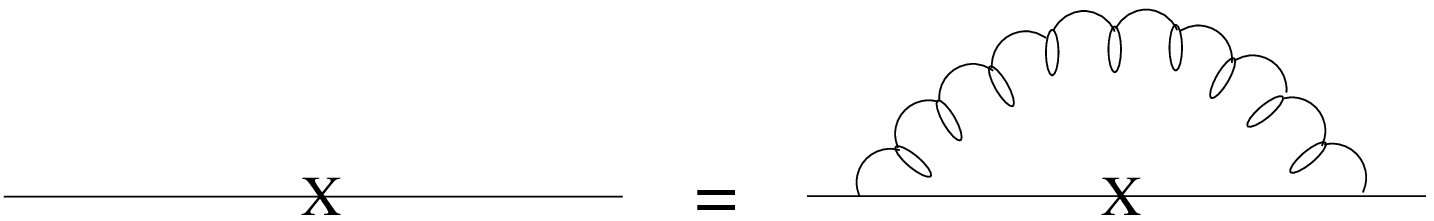}
\end{center}
\vspace*{-1cm}
\caption{Graphical form of the self-consistent equation for the
condensate (gap equation).}
\end{figure}


With a contact interaction, and throwing away manifestly spurious
ultraviolet divergences, we obtain a gap equation of the type
\begin{equation}
\Delta \propto  g^2 \int d\epsilon {\Delta \over \sqrt {\epsilon^2 +
\Delta^2}}. 
\end{equation}
The phase space transverse to the Fermi surface cancels against a
propagator, leaving the integral over the longitudinal distance
$\epsilon$ to the Fermi surface.  Note that the integral on the right
diverges at small $\epsilon$, so that as long as the proportionality
constant is positive one will have non-trivial solutions for $\Delta$,
no matter how small is $g$.  Indeed, one finds that for small $g$,
$\Delta \sim e^{-{\rm const}/g^2}$.

If we restore the gluon propagator, we will find a non-trivial angular
integral, which diverges for forward scattering.  That divergence will
be killed, however, if the gluon acquires a mass $\propto g\Delta$
through the Meissner-Higgs mechanism.  Thus we arrive at a gap
equation of the type
\begin{equation}
\Delta \propto g^2 \int d\epsilon {\Delta \over \sqrt {\epsilon^2 +
\Delta^2}} dz { \mu^2 \over \mu^2 + (g\Delta )^2 }. 
\end{equation}
Now one finds $\Delta \sim e^{-{\rm const}/g}$!

A proper discussion of the microscopic gap equation is considerably
more involved than this, but the conclusion that the gap goes
exponentially in the inverse coupling (rather than its square) at weak
coupling still emerges.  It has the amusing consequence, that at
asymptotically high densities the gap becomes arbitrarily large!
This is because asymptotic freedom insures that it is the microscopic
coupling $1/g(\mu)^2$ which vanishes logarithmically, so that
$e^{-{\rm const}/g(\mu ) }$ does not shrink as fast as $1/\mu$.  Since
the ``dimensional analysis'' scale of the gap is set by $\mu$,
its linear growth wins out asymptotically.

\section{Lecture 5: High-Density QCD: Color-Flavor Locking and Quark-Hadron 
Continuity}

Color-flavor locking has many remarkable consequences.
There is a gap for all colored excitations, including the gluons.
This is, operationally, confinement.  The photon picks up a gluonic
component of just such a form as to ensure that all elementary
excitations, including quarks, are integrally charged.  Some of the
gluons acquire non-zero, but integer-valued, electric charges.  Baryon
number is spontaneously broken, which renders the high-density
material a superfluid.

If in addition the quarks are massless, then their chiral symmetry is
spontaneously broken, by a new mechanism.  The left-dynamics and the
right-dynamics separately lock to color; but since color allows only
vector transformations, left is thereby locked to right.

You may notice several points of resemblance between the low-energy
properties calculated for the high-density color-flavor locked phase
and the ones you might expect at low density, based on
semi-phenomenological considerations such as the MIT bag model, or
experimental results in real-world QCD.  The quarks play the role of
low-lying baryons, the gluons play the role of the low-lying vector
mesons, and the Nambu-Goldstone bosons of broken chiral symmetry play
the role of the pseudoscalar octet.  All the quantum numbers match,
and the spectrum has gaps -- or not -- in all the right places.  In
addition we have baryon number superfluidity, which extends the
expected pairing phenomena in nuclei.  Overall, there is an uncanny
match between all the universal, and several of the non-universal,
features of the calculable high-density and the expected low-density
phase.  This leads us to suspect that there is no phase transition
between them!

\subsection{Gauge Symmetry (non)-Breaking}

An aspect of Eqn.~\ref{Cond} that might appear troubling at first
sight, is its lack of gauge invariance.  There are powerful general
arguments that local gauge invariance cannot be broken.
Indeed, local gauge invariance is really a tautology, stating the
equality between redundant variables.  Yet its `breaking' is
central to two of the most successful theories in physics, to wit BCS
superconductivity theory and the standard model of electroweak
interactions.  In BCS theory we postulate a non-zero vacuum
expectation value for the (electrically charged) Cooper pair field,
and in the standard model we postulate a non-zero vacuum expectation
value for the Higgs field, which violates both the weak isospin SU(2)
and the weak hypercharge U(1).

In each case, we should interpret the condensate as follows.  We are
working in a gauge theory at weak coupling.  It is then very
convenient to fix a gauge, because after we have done so -- but not
before! -- the gauge potentials will make only small fluctuations
around zero, which we will be able to take into account
perturbatively.  Of course at the end of any calculation we must
restore the gauge symmetry, by averaging over the gauge fixing
parameters (gauge unfixing).  Only gauge-invariant results will
survive this averaging.  In a fixed gauge, however, one might capture
important correlations, that characterize the ground state, by
specifying the existence of non-zero condensates relative to that
gauge choice.  These condensates need not, and generally will not,
break any symmetries.

For example, in the standard electroweak model one employs a non-zero
vacuum expectation value for a Higgs doublet field $\langle \phi^a
\rangle ~=~ v \delta^a_1$, which is not gauge invariant.  One might be
tempted to use the magnitude of its absolute square, which is gauge
invariant, as an order parameter for the symmetry breaking, but
$\langle \phi ^\dagger \phi \rangle $ never vanishes, whether or not
any symmetry is broken (and, of course, $\langle \phi ^\dagger \phi
\rangle $ breaks no symmetry).  In fact there is no order parameter
for the electroweak phase transition, and it has long been appreciated
that one could, by allowing the $SU(2)$ gauge couplings
to become large, go over into a `confined' regime while
encountering no sharp phase transition.  The most important
gauge-invariant consequences one ordinarily infers from the
condensate, of course, are the non-vanishing W and Z boson masses.
This absence of massless bosons and long-range forces is the essence
of confinement, or of the Meissner-Higgs effect.  Evidently, when used
with care, the notion of spontaneous gauge symmetry breaking can be an
extremely convenient fiction -- so it proves for Eqn.~\ref{Cond}.

\subsection{Symmetry Accounting}

The equations of our original model, QCD with three massless flavors,
has the continuous symmetry group $SU(3)^c \times SU(3)_L \times
SU(3)_R \times U(1)_B$.  The Kronecker deltas that appear in the
condensate Eqn.~\ref{Cond} are invariant under neither color nor
left-handed flavor nor right-handed flavor rotations separately. Only
a global, diagonal $SU(3)$ leaves the ground state invariant. Thus we
have the symmetry breaking pattern
\begin{equation}
SU(3)^c \times SU(3)_L \times SU(3)_R \times U(1)_B \rightarrow
SU(3)_{c+L+R} \times Z_2~. 
\end{equation}

\subsubsection{Confinement}

The breaking of local color symmetry implies that all the gluons
acquire mass, according to the Meissner (or alternatively Higgs)
effect.  There are no long-range, $1/r$ interactions.  There is no
direct signature for the color degree of freedom -- although, of
course, in weak coupling one clearly perceives its avatars.  It is
veiled or, if you like, confined.

\subsubsection{Chiral Symmetry Breaking}

If we make a left-handed chiral rotation then we must compensate it
by a color rotation, in order to leave the left-handed condensate invariant.
Color rotations being vectorial, we must then in addition make a
right-handed chiral rotation, in order to leave the right-handed
condensate invariant.  Thus chiral symmetry is spontaneously broken,
by a new mechanism: although the left- and right- condensates are
quite separate (and, before we include instantons -- see below -- not
even phase coherent), because both are locked to color they are
thereby locked to one another.

The spontaneous breaking of global chiral $SU(3)_L \times SU(3)_R$
brings with it an octet of pseudoscalar Nambu-Goldstone bosons,
collective modes interpolating, in space-time, among the condensates
related by the lost symmetry.  These massless modes, as is familiar,
are derivatively coupled, and therefore they do not generate singular
long-range interactions.

\subsubsection{Superfluidity and `Nuclear' Pairing}

Less familiar, and perhaps disconcerting at first sight, is the loss
of baryon number symmetry.  This does not, however, portend proton
decay, any more than does the non-vanishing condensate of helium atoms
in superfluid ${\rm He4}$.  Given an isolated finite sample, the current
divergence equation can be integrated over a surface surrounding the
sample, and unambiguously indicates overall number conservation.  To
respect it, one should project onto states with a definite number of
baryons, by integrating over states with different values of the
condensate phase.  This does not substantially alter the physics of
the condensate, however, because the overlap between states of
different phase is very small for a macroscopic sample.  Roughly
speaking, there is a finite mismatch per unit volume, so the overlap
vanishes exponentially in the limit of infinite volume.  The true
meaning of the formal baryon number violation is that there are
low-energy states with different distributions of baryon number, and
easy transport among them.  Indeed, the dynamics of the condensate is
the dynamics of superfluidity:  gradients in the Nambu-Goldstone mode
are none other than the superfluid flow.

We know from experience that large nuclei exhibit strong even-odd
effects, and an extensive phenomenology has been built up around the
idea of pairing in nuclei.  If electromagnetic Coulomb forces
didn't spoil the fun, we could confidently expect that extended
nuclear matter would exhibit the classic signatures of superfluidity.
In our 3-flavor version the Coulomb forces do not come powerfully into
play, since the charges of the quarks average out to electric
neutrality.  Furthermore, the tendency to superfluidity exhibited by
ordinary nuclear matter should be enhanced by the additional channels
operating coherently.  So one should expect strong superfluidity at
ordinary nuclear density, and it becomes less surprising that we find
it at asymptotically large density too.

\subsubsection{Global Order Parameters}

I mentioned before that the Higgs mechanism as it operates in the
electroweak sector of the standard model has no gauge-invariant
signature.  With color-flavor locking we're in better shape,
because global as well as gauge symmetries are broken.  Thus there are
sharp differences between the color-flavor locked phase and the free
phase. There must be phase transitions -- as a function, say, of
temperature -- separating them.

In fact, it is a simple matter to extract gauge invariant order
parameters from our primary, gauge variant condensate at weak
coupling.  For instance, to form a gauge invariant order parameter
capturing chiral symmetry breaking we may take the product of the
left-handed version of Eqn.~\ref{Cond} with the right-handed version
and saturate the color indices, to obtain
\begin{equation}
\label{chiral}
\langle q^\alpha_{La} q^\beta_{Lb} \bar q^c_{R\alpha} \bar q^d_{R\beta} \rangle
\sim \langle q^\alpha_{La} q^\beta_{Lb}\rangle \langle \bar
q^c_{R\alpha} \bar q^d_{R\beta} \rangle \sim (\kappa_1^2 + \kappa_2^2)
\delta^c_a \delta^d_b  
+ 2 \kappa_1 \kappa_2 \delta^d_a \delta^c_b
\end{equation}
Likewise we can take a product of three copies of the condensate and
saturate the color indices, to obtain a gauge invariant order
parameter for superfluidity.  These secondary order parameters will
survive gauge unfixing unscathed.  Unlike the primary condensate from
which they were derived, they are not just convenient fictions, but 
measurable realities.

\subsubsection{A Subtlety: Axial Baryon Number}

As it stands the chiral order parameter Eqn.~\ref{chiral} is not quite the
usual one, but roughly speaking its square.  It leaves invariant an
additional $Z_2$, under which the left-handed quark fields change
sign.  Actually this $Z_2$ is not a legitimate symmetry of the full
theory, but suffers from an anomaly.

Since we can be working at weak coupling, we can be more specific.
Our model Hamiltonian Eqn.~\ref{modelH} was abstracted from one-gluon
exchange, which is the main interaction among high-energy quarks in
general, and so in particular for modes near our large Fermi surfaces.
The instanton interaction is much less important, at least
asymptotically, both because it is intrinsically smaller for energetic
quarks, and because it involves six fermion fields, and hence (one can
show) is irrelevant as one renormalizes toward the Fermi surface.
However, it represents the leading contribution to axial baryon number
violation.  In particular, it is only $U_A(1)$ violating interactions
that fix the relative phase of our left- and right- handed
condensates.  So a model Hamiltonian that neglects them will have an
additional symmetry that is not present in the full theory.  After
spontaneous breaking, which does occur in the axial baryon number
channel, there will be a Nambu-Goldstone boson in the model theory,
that in the full theory acquires an anomalously (pun intended)
small mass.  Similarly, in the full theory there will
be a non-zero tertiary chiral condensate of the usual kind, bilinear
in quark fields, but it will be parametrically smaller than
Eqn.~\ref{chiral}.

\subsection{Elementary Excitations}

There are three sorts of elementary excitations.  They are the modes
produced directly by the fundamental quark and gluon fields, and the
collective modes connected with spontaneous symmetry breaking.

The quark fields of course produce spin $1/2$ fermions.  Some of these
are true long-lived quasiparticles, since there is nothing for them to
decay into.  They form an octet and a singlet under the residual
diagonal $SU(3)$.  There is an energy gap for production of pairs
above the ground state.  Actually there are two gaps: a smaller one
for the octet, and a larger one for the singlet.

The gluon fields produce an octet of spin $1$ bosons.  As previously
mentioned, they acquire a mass by the Meissner-Higgs phenomenon.  We
have already discussed the Nambu-Goldstone bosons, too.

\subsection{A Modified Photon}

The notion of `confinement' I advertised earlier, phrased in terms
of mass gaps and derivative interactions, might seem rather
disembodied.  So it is interesting to ask whether and how a more
traditional and intuitive criterion of confinement -- no fractionally
charged excitations -- is satisfied.

Before discussing electromagnetic charge we must identify the unbroken
gauge symmetry, whose gauge boson defines the physical photon in our
dense medium.  The original electromagnetic gauge invariance is
broken, but there is a combination of the original electromagnetic
gauge symmetry and a color transformation which leaves the condensate
invariant.  Specifically, the original photon $\gamma$ couples
according to the matrix
\begin{equation}
\left( \begin{array}{ccc}
{2 \over 3} & 0 & 0 \\
0& -{1\over 3} & 0 \\
0& 0& -{1\over 3}
\end{array} \right)
\end{equation}
in flavor space, with strength $e$.  There is a gluon $G$ which
couples to the matrix
\begin{equation}
\left( \begin{array}{ccc}
-{2 \over 3} & 0 & 0 \\
0& {1\over 3} & 0 \\
0& 0& {1\over 3}
\end{array} \right)
\end{equation}
in color space, with strength $g$.   Then the combination 
\begin{equation}
\label{photon}
\tilde \gamma = {g \gamma + e G \over \sqrt{e^2 + g^2}}
\end{equation}
leaves the `locking' Kronecker deltas in color-flavor space
invariant.  In our medium, it represents the physical photon.  What
happens here is similar to what occurs in the electroweak sector of
the standard model, where both weak isospin and weak hypercharge are
separately broken by the Higgs doublet, but a cunning combination
remains unbroken, and defines electromagnetism.

\subsubsection{Integer Charges!}

Now with respect to $\tilde \gamma$ the electron charge is
\begin{equation}
{-eg\over \sqrt {e^2 + g^2}}~,
\end{equation}
deriving of course solely from the $\gamma$ piece of Eqn.~\ref{photon}.
The quarks have one flavor and one color index, so they pick up
contributions from  both pieces.  In each sector we find the
normalized charge unit ${eg\over \sqrt {e^2 + g^2}}$, and it is
multiplied by some choice from among $(2/3, -1/3, -1/3)$ or $(-2/3,
1/3, 1/3)$ respectively.  The total, obviously, can be $\pm 1$ or $0$.
Thus the excitations produced by the quark fields are integrally
charged, in units of the electron charge.  Similarly the gluons have
an upper color and a lower anti-color index, so that one faces similar
choices, and reaches a similar conclusion.  In particular, some of the
gluons have become electrically charged.  The pseudoscalar
Nambu-Goldstone modes have an upper flavor and a lower anti-flavor
index, and yet again the same conclusions follow.  The superfluid
mode, of course, is electrically neutral.

\subsubsection{A Conceptual Jewel}

It is fun to consider how a chunk of our color-flavor locked material
would look.  If the quarks were truly massless, then so would be
Nambu-Goldstone bosons (at the level of pure QCD), and one might
expect a rather unusual `bosonic metal', in which low-energy
electromagnetic response is dominated by these modes.  Actually
electromagnetic radiative corrections lift the mass of the charged
Nambu-Goldstone bosons, creating a gap for the charged channel.  The
same effect would be achieved by turning on a common non-zero quark
mass.  Thus the color-flavor locked material forms a transparent
insulator.  Altogether it would resemble a diamond, reflecting portions
of incident light waves, but allowing finite portions through and out
again!

\subsection{Quark-Hadron Continuity}

The universal features of the color-flavor locked state: confinement,
chiral symmetry breaking down to vector $SU(3)$, and superfluidity,
are just what one would expect, based on standard phenomenological
models and experience with real-world QCD at low density.  Now we see
that the low-lying spectrum likewise bears an uncanny resemblance to
what one finds in the Particle Data Book (or rather what one would
find, in a world of three degenerate quarks).  It is hard to resist
the inference that there is no phase transition separating them.  Thus
there need not be, and presumably is not, a sharp distinction between
the low-density phase, where microscopic calculations are difficult
but the convenient degrees of freedom are ``obviously'' hadrons,
and the asymptotic high-density phase, where weak-coupling (but
non-perturbative) calculations are possible, and the right degrees of
freedom are elementary quarks and gluons plus collective modes
associated with spontaneous symmetry breaking.  We call this
quark-hadron continuity.  It might seem shocking that a
quark can ``be'' a baryon, but remember that it is immersed in a
sea of diquark condensate, wherein the distinction between one quark
and three becomes negotiable.

\subsection{Remembrance of Things Past}

An entertaining aspect of the emergent structure is that two beautiful
ideas from the pre-history of QCD, that were bypassed in its later
development, have come very much back to center stage, now with
microscopic validation.  The quark-baryons of the color-flavor locked
phase follow the charge assignments proposed by Han and Nambu.  And
the gluon-vector mesons derive from the Yang-Mills 
gauge principle -- as originally proposed, for rho mesons!

\subsection{More Quarks}

For larger numbers of quarks, the story is qualitatively similar.
Color symmetry is broken completely, and there is a gap 
in all quark channels, so the weak-coupling treatment is adequate.
Color-flavor locking is so favorable that there seems to be a
periodicity: if the number of quarks is a multiple of three, one finds
condensation into 3$\times$3 blocks, while if it is 4+3k or 5+3k one
finds k color-flavor locking blocks together with special patterns
characteristic of 4 or 5 flavors.

There is an amusing point here.  QCD with a very large number of
massless quarks, say 16, has an infrared fixed point at very weak
coupling.  Thus it should be quasi-free at
zero density, forming a nonabelian Coulomb phase, featuring conformal
symmetry, no confinement, and no chiral symmetry breaking.  To say the
least, it does not much resemble real-world QCD.  There are
indications that this qualitative behavior may persist even for
considerably fewer quarks (the critical number might be as small as 5
or 6).  Nevertheless, at high density, we have discovered, these
many-quark theories all support more-or-less normal-looking `nuclear
matter' -- including confinement and chiral symmetry breaking!

\subsection{Fewer Quarks, and Reality}

One can perform a similar analysis for two quark flavors.  A new
feature is that the instanton 
interaction now involves four rather than six quark legs, so it
remains relevant as one renormalizes toward the Fermi surface.  Either
the one-gluon exchange or the instanton interaction, treated in the
spirit above, favors condensation of the form
\begin{equation}
\label{twoCond}
\begin{array}{lll}
~\langle q^{i\alpha}_{La} (p) q^{j\beta}_{Lb} (-p) \rangle ~=~ 
-\langle q^{i\alpha}_{Ra} (p) q^{j\beta}_{Lb} (-p) \rangle ~=~ 
\epsilon^{ij} \kappa(p^2) \epsilon^{\alpha\beta3} \epsilon_{ab}~.
\end{array}
\end{equation}
Formally, Eqn.~\ref{twoCond} is quite closely related to Eqn.~\ref{chiral},
since $\epsilon^{\alpha\beta I} \epsilon_{abI} = 2(\delta^\alpha_a
\delta^\beta_b - \delta^\alpha_b \delta^\beta_a)$.  Their physical
implications, however, are quite different.

To begin with, Eqn.~\ref{twoCond} does not lead to gaps in all quark
channels.  The quarks with color labels 1 and 2 acquire a gap, but
quarks of the third color of quark are left untouched.  Secondly, the
color symmetry is not completely broken.  A residual $SU(2)$, acting
among the first two colors, remains valid.  For these reasons,
perturbation theory about the ground state defined by Eqn.~\ref{twoCond} is
{\it not\/} free of infrared divergences, and we do not have a fully
reliable grip on the physics.

Nevertheless it is plausible that the qualitative features suggested
by Eqn.~\ref{twoCond} are not grossly misleading.  The residual $SU(2)$
presumably produces confined glueballs of large mass, and assuming
this occurs, the residual gapless quarks are weakly coupled.

Assuming for the moment that no further condensation occurs, for
massless quarks we have the symmetry breaking pattern
\begin{equation}
SU(3)^c \times SU(2)_L \times SU(2)_R \times U(1)_B \rightarrow SU(2)^c \times 
SU(2)_L \times SU(2)_R \times {\tilde U}(1)_B
\end{equation}
Here the modified baryon number acts only on the third color of
quarks.  It is a combination of the original baryon number and a color
generator, that are separately broken but when applied together leave
the condensate invariant.  Comparing to the zero-density ground state,
one sees that color symmetry is reduced, chiral symmetry is restored,
and baryon number is modified.  Only the restoration of chiral
symmetry is associated with a legitimate order parameter, and only it
requires a sharp phase transition.

In the real world, with the $u$ and $d$ quarks light but not strictly
massless, there is no rigorous argument that a phase transition is
necessary.  It is (barely) conceivable that one might extend
quark-hadron continuity to this case.  Due to medium
modifications of baryon number and electromagnetic charge the
third-color $u$ and $d$ quarks have the quantum numbers of nucleons.
The idea that chiral symmetry is effectively restored in nuclear
matter, however, seems problematic quantitatively.  More plausible,
perhaps, is that there is a first-order transition between nuclear
matter and quark matter.  This is suggested by some model
calculations, and is the basis for an attractive 
interpretation of the MIT bag model, according to which baryons are
droplets wherein chiral symmetry is restored.

\subsubsection{Thresholds and Mismatches}
In the real world there are two quarks, $u$ and $d$, whose mass is
much less than $\Lambda_{QCD}$, and one, $s$, whose mass is comparable
to it.  Two simple qualitative effects, that have major implications
for the zero-temperature phase diagram, arise as consequences of this
asymmetric spectrum.  They are expected,
whether one analyzes from the quark side or from the hadron side.

The first is that one can expect a threshold, in chemical potential
(or pressure), for the appearance of any strangeness at all in the
ground state.  This will certainly hold true in the limit of large
strange quark mass, and there is considerable evidence for it in the
real world.  This threshold is in addition to the threshold
transitions at lower chemical potentials, from void to nuclear matter,
and (presumably) from nuclear matter to two-flavor quark matter, as
discussed above.

The second is that at equal chemical potential the Fermi surfaces of
the different quarks will not match.  This mismatch cuts off the
Cooper instability in mixed channels.  If the nominal gap is large
compared to the mismatch, one can treat the mismatch as a
perturbation.  This will always be valid at asymptotically high
densities, since the mismatch goes as $m_s^2/\mu$, whereas the gap
eventually grows with $\mu$.  If the nominal gap is small compared to
the mismatch, condensation will not occur.

\subsubsection{Assembling the Pieces}
With these complications in mind, we can identify three major phases
in the plane of chemical potential and strange quark mass, that
reflect the simple microscopic physics we have surveyed above.  (There
might of course be additional ``minor'' phases -- notably
including normal nuclear matter!)  There is 2-flavor quark matter,
with restoration of chiral symmetry, and zero strangeness.  Then there
is a 2+1-flavor phase, in which the strange and non-strange Fermi
surfaces are badly mismatched, and one has independent dynamics for
the corresponding low energy excitations.  Here one expects
strangeness to break spontaneously, by its own Fermi surface
instability.  Finally there is the color-flavor locked phase. A
caricature version of the phase diagram in the $(m_s, \mu)$ plane,
illustrating these features, is given in Figure 13.


\begin{figure}[htb]
\centerline{\psfig{figure=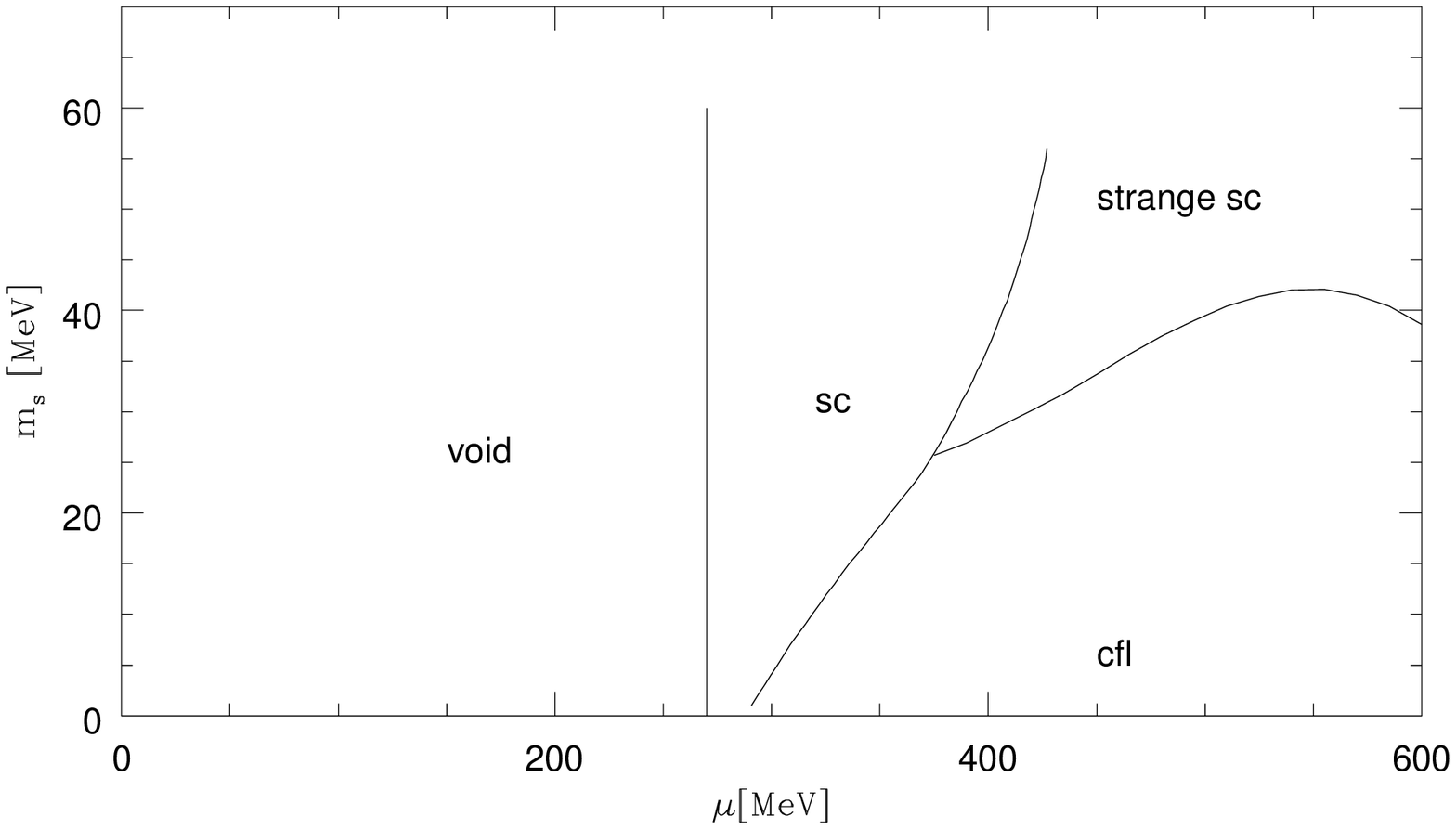,width=10cm}}
\vglue-.4in
\caption[]{Phase diagram calculated in a schematic model.}
\label{fig:phase.ps}
\end{figure}

\subsubsection{Reality}

The progress reported here, while remarkable, mainly concerns the asymptotic
behavior of high-density QCD.  Its extrapolation to practical densities is at
present semi-quantitative at best.  To do real justice to the
potential applications, we need to learn how to do more accurate
analytical and numerical work at moderate densities.

As regards analytical work, we can take heart from some recent
progress on the equation of state at high temperature.  Here there are
extensive, interesting numerical 
results, which indicate that the behavior is quasi-free,
but that there are very significant quantitative corrections to free
quark-gluon plasma results, especially for the pressure.  Thus it is
plausible {\it a priori\/} that some weak-coupling, but
non-perturbative, approach will be workable, and this seems to be
proving out.  An encouraging feature here is that the analytical
techniques used for high temperature appear to be capable of extension
to finite density without great difficulty.

Numerical work at finite density, unfortunately, is plagued by poor
convergence.  This arises because the functional integral is not
positive definite configuration by configuration, so that importance
sampling fails, and one is left looking for a small residual from much
larger canceling quantities.

There are cases in which this problem does not arise.  It does not
arise for two colors.  Although low-density hadronic
matter is quite different in a two-color world than a three-color
world -- the baryons are bosons, so one does not get anything like a
shell structure for nuclei -- I see no reason to expect that the
asymptotic, high-density phases should be markedly different.  It
would be quite interesting to see Fermi-surface behavior arising for
two colors at high density (especially, for the ground state
pressure), and even more interesting to see the effect of diquark
condensations.

Another possibility, that I have been discussing with David Kaplan, is
to engineer lattice gauge theories whose low-energy excitations
resemble those of finite density QCD near the Fermi surface, but which
are embedded in a theory that is globally particle-hole symmetric, and
so feature a positive-definite functional integral.

Aside from these tough quantitative issues, there are a number of
directions in which the existing work should be expanded and
generalized, that appear to be quite accessible.  There is already a
rich and important theory of the behavior of QCD at non-zero
temperature and zero baryon number density.  We should construct a
unified picture of the phase structure as a function of both
temperature and density; to make it fully illuminating, we should also
allow at least the strange quark mass to vary.  We should allow for
the effect of electromagnetism (after all, this is largely what makes
neutron stars what they are) and of rotation.  We should consider
other possibilities than a common chemical potential for all the
quarks.

As physicists we should not, however, be satisfied with hoarding up
formal, abstract knowledge.  There are concrete experimental
situations and astrophysical objects we must speak to.  Hopefully,
having mastered some of the basic vocabulary and grammar, we will soon
be in a better position to participate in a two-way dialogue with
Nature.

\newpage

{\bf Background Material}
\bigskip

\begin{itemize}

\item  General background on quantum field theory and QCD:  

\begin{enumerate}

\item T.-P. Cheng and L.-F. Li, {\it Gauge Theory of Elementary Particles\/}
(Oxford University Press, London 1984),

\item M. Peskin and D. Schroder, {\it Introduction to Quantum Field
Theory\/} (Addison-Wesley, Redwood City California, 1995).
\end{enumerate}

\item Lattice gauge theory:

\begin{enumerate}

\item M. Creutz, {\it Quarks, Gluons, and Lattices\/} (Cambridge University
Press, Cambridge U.K., 1983).

\end{enumerate}

\item Chiral symmetry and current algebra:

\begin{enumerate}

\item  S. Weinberg, {\it Quantum Theory of Fields II\/} chapter 19  
(Cambridge University
Press, Cambridge U.K., 1996).

\end{enumerate}

\item Renormalization group for critical phenomena:

\begin{enumerate}

\item D. Amit, {\it Field Theory, the Renormalization Group, and Critical
Phenomena\/} (World Scientific, Singapore 1984).
\end{enumerate}

\item Renormalization group toward the Fermi surface:

\begin{enumerate}

\item J. Polchinski, hep-th/9210046, 

\item R. Shankar, {\it Rev. Mod. Phys}. {\bf 66}, 129 (1993).

\end{enumerate}

\item Superconductivity:

\begin{enumerate}

\item J. R. Schrieffer, {\it Theory of Superconductivity\/}
(Benjamin-Cummins, Reading Mass., 1984).

\end{enumerate}

\item Tricritical points:

\begin{enumerate}
\item I. Lawrie and S. Sarbach, in {\it Phase Transitions and Critical
Phenomena 9\/} eds. C. Domb and J. Lebowitz (Academic, New York 1984).
\end{enumerate}

\item Quark-gluon plasma:

\begin{enumerate}

\item U. Heniz and M. Jacob, nucl-th/0002042.
\end{enumerate}

\newpage

{\bf  Sources and Further Reading}
\bigskip

\item The perspective on QCD adopted in Lecture 1, as a series of
interweaving stories of symmetry and dynamics, is taken from my
monograph {\it QCD\/} in preparation for Princeton University Press.

\item The material on high-temperature phase transitions in Lectures 2 and 3
is mostly taken from 

\begin{enumerate}
\item F. Wilczek, {\it Int. Jour. Mod. Phys.} {\bf A7} 3911, (1992).

\item K. Rajagopal and F. Wilczek, {\it Nucl. Phys.} {\bf B399)} 395, (1993).
\end{enumerate}

\item The possibility of a true critical point was pointed out in 
M. Stephanov, K. Rajagopal, and E. Shuryak {\it Phys. Rev. Lett}. {\bf
81}, 4816 (1998).

\item Its possible experimental signatures are discussed in 
  M. Stephanov, hep-ph/9906242.

\item For recent lattice results see F. Karsch, hep-lat/9909006.

\item Lectures 4 and 5 are based on F. Wilczek, hep-ph/9908480.

\item For further developments see especially
T. Sch\"afer, hep-ph/9909574.

\end{itemize}
These papers contain numerous additional references.  Obviously,
several topics discussed in the Lectures are under very active
development, and you should consult the web for up-to-date results.

\end{document}